\documentclass[11pt,fleqn]{article}

\usepackage{a4}
\usepackage{amstext}
\usepackage{amsfonts}
\usepackage{amssymb}
\usepackage{booktabs}
\usepackage{color}
\usepackage{caption}         
\usepackage{cite}
\usepackage{dsfont}
\usepackage{epsfig}
\usepackage{mathtools}
\usepackage{multirow}
\usepackage{physics}
\usepackage{ulem}
\usepackage{subcaption}

\newcommand{\gtapprox}{\raisebox{-0.5ex}{$\,\stackrel{>}{\scriptstyle\sim}\,$}}
\newcommand{\ltapprox}{\raisebox{-0.5ex}{$\,\stackrel{<}{\scriptstyle\sim}\,$}}

\begin{document}


\begin{center}

{\huge \bf Precision computation of hybrid static}

{\huge \bf potentials in SU(3) lattice gauge theory}

\vspace{0.5cm}

\textbf{Stefano Capitani, Owe Philipsen, Christian Reisinger, Carolin Riehl, Marc Wagner}

Goethe-Universit\"at Frankfurt am Main, Institut f\"ur Theoretische Physik, Max-von-Laue-Stra{\ss}e 1, D-60438 Frankfurt am Main, Germany

\vspace{0.5cm}

November 27, 2018

\end{center}

\begin{tabular*}{16cm}{l@{\extracolsep{\fill}}r} \hline \end{tabular*}

\vspace{-0.4cm}
\begin{center} \textbf{Abstract} \end{center}
\vspace{-0.4cm}

We perform a precision computation of hybrid static potentials with quantum numbers $\Lambda_\eta^\epsilon = \Sigma_g^-,\Sigma_u^+,\Sigma_u^-,\Pi_g,\Pi_u,\Delta_g,\Delta_u$ using SU(3) lattice gauge theory. The resulting potentials are used to estimate masses of heavy $\bar{c} c$ and $\bar{b} b$ hybrid mesons in the Born-Oppenheimer approximation. Part of the lattice gauge theory computation, which we discuss in detail, is an extensive optimization of hybrid static potential creation operators. The resulting optimized operators are expected to be essential for future projects concerning the computation of 3-point functions as e.g.\ needed to study spin corrections, decays or the gluon distribution of heavy hybrid mesons.

\begin{tabular*}{16cm}{l@{\extracolsep{\fill}}r} \hline \end{tabular*}

\thispagestyle{empty}


\newpage

\setcounter{page}{1}

\section{Introduction}

The success of the quark model, following the realization of the importance of $\textrm{SU(3)}$ flavor symmetry in the context of the eightfold way, led to understanding the properties of a large number of mesons and baryons. However, the quark model does not contain gluons. In the framework of QCD it is, thus, of utmost importance to investigate and to understand what kind of additional hadronic states or resonances can appear, when gluons are allowed to be in excited states. In this work we are particularly interested in heavy hybrid mesons, i.e.\ in mesons composed of heavy $c$ or $b$ quarks, where gluons contribute to the quantum numbers $J^{P C}$ in a non-trivial way.

With the discovery of the first of the so-called $XYZ$ mesons around fifteen years ago, the $X(3872)$, an entirely new chapter of hadronic physics was opened and flourished. At present there are about thirty such exotic states, which have been observed (for a theoretical summary cf.\ e.g.\ \cite{Braaten:2014ita,Meyer:2015eta,Swanson:2015wgq,Lebed:2016hpi}, for an experimental review cf.\ e.g.\ \cite{Olsen:2017bmm}). Many of these exotic states are believed to be tetraquark resonances, but some of them are also being considered as candidates for hybrid mesons. It is very challenging to understand the internal structure of exotic hadrons and, even though there is little doubt that hybrid mesons and baryons exist, not much else is known about them.

It is a notable feature of hybrid mesons that, due to their excited gluonic degrees of freedom, part of them have $J^{P C}$ quantum numbers, which are forbidden in the quark model \footnote{In the quark model $P = (-1)^{L+1}$ and $C = (-1)^{L+S}$ for a meson, where $L \in \{ 0,1,2,\ldots \}$ is the orbital angular momentum and $S \in \{ 0,1 \}$ is the quark spin.}. In this sense, observing a meson with $J^{P C} = 0^{+ -} , 0^{- -} , 1^{- +} , 2^{+ -}, \ldots$ indicates an exotic structure, possibly the presence of excited gluons.

Obtaining solid results for exotic hadrons is highly non trivial, both on the theoretical and on the experimental side. There is currently a lot of experimental activity in the field of exotic hadrons, but even the exact attribution of the $J^{P C}$ quantum numbers is difficult for many of the experimentally observed $XYZ$ states. Moreover, all up to now experimentally observed heavy candidates for hybrid mesons exhibit non-exotic quantum numbers, which makes theoretical investigations of their properties even more important. There are a few exotic states, which could be heavy hybrid mesons, the most prominent candidate being the $Y(4260)$ with quantum numbers $J^{P C}=1^{- -}$, but there are also arguments disfavoring a hybrid identification (cf.\ e.g.\ the discussions in \cite{Braaten:2014qka,Berwein:2015vca}). Several existing and future experiments will be taking data in the next couple of years (for example the GlueX and the PANDA experiment) and, thus, many more candidates for hybrid mesons are likely to be discovered. On the theoretical side models devised to explain the properties of exotic hadrons typically possess a limited applicability and there is no overall coherent theoretical picture of these hadrons. Lattice field theory, however, which is a non-perturbative first principles approach, is an ideal method to predict masses or to explore properties of hybrid mesons. Such results might also be useful as input for effective theories like pNRQCD or to calibrate or devise improved theoretical models.

In this work we carry out a precise computation of several hybrid static potentials using SU(3) lattice gauge. Gluonic excitations are included by considering trial states containing a static quark-antiquark pair and gluons, which are characterized by non-trivial quantum numbers, i.e.\ orbital angular momentum, parity or charge conjugation. Our aim is to improve on existing similar lattice field theory computations \cite{Griffiths:1983ah,Campbell:1984fe,Campbell:1987nv,
Michael:1990az,Perantonis:1990dy,Juge:1997nc,Peardon:1997jr,
Juge:1997ir,Morningstar:1998xh,Michael:1998tr,
Juge:1999ie,Juge:1999aw,Michael:1999ge,Bali:2000vr,
Morningstar:2001nu,Juge:2002br,Michael:2003ai,Juge:2003qd,
Michael:2003xg,Bali:2003jq,Juge:2003ge,
Wolf:2014tta,Reisinger:2017btr,Bicudo:2018yhk,Bicudo:2018jbb,Reisinger:2018lne}) by providing results with smaller statistical errors and at finer spatial resolution (section~\ref{SEC499}) and by discussing all technical details of the optimization of creation operators and the computation of the potentials (section~\ref{sec:1} to section~\ref{SEC445}). The latter is a necessary and important preparatory step for the computation of 3-point functions, which we recently started, and which we briefly discuss in our conclusions in section~\ref{SEC602}.

We also use some of the resulting hybrid static potentials to estimate masses of heavy-quark hybrid mesons, where the quarks are either $\bar{c} c$ or $\bar{b} b$ (section~\ref{SEC498}). This is done in the Born-Oppenheimer approximation \cite{bo}, where effects from the quark spins are neglected, by numerically solving an appropriate Schr\"odinger equation. This is expected to be a good approximation, because the time scales of the gluons and of the heavy charm or bottom quarks are significantly different and, thus, their dynamics decouples almost completely. In this context also effective theory approaches like potential Non Relativistic QCD (pNRQCD) are extremely useful, for example when parameterizing discrete lattice field theory results for hybrid static potentials by continuous functions (for recent pNRQCD articles on hybrid mesons cf.\ e.g. \cite{Berwein:2015vca,Brambilla:2017uyf,Brambilla:2018pyn}).


\newpage

\section{\label{sec:1}Quantum numbers and trial states}

A hybrid static potential is a potential of a static quark and a static antiquark, where the gluons form non-trivial structures and contribute to the quantum numbers. We compute such hybrid static potentials from Wilson loop-like correlation functions using SU(3) lattice gauge theory. The gluonic excitations are realized by replacing the straight spatial Wilson lines of the Wilson loops by parallel transporters, which have a less trivial structure.

We put the static quark and the static antiquark, which we treat as spinless color charges, at positions $\mathbf{r}_Q=(0,0,+r/2)$ and $\mathbf{r}_{\bar{Q}}=(0,0,-r/2)$, respectively, i.e.\ separate them along the $z$ axis. In the following we omit the $x$ and the $y$ coordinate, e.g.\ $Q(+r/2) \equiv Q(0,0,+r/2)$.

Hybrid static potentials are characterized by the following quantum numbers:
\begin{itemize}
\item $\Lambda = 0,1,2,\ldots$, the absolute value of the total angular momentum with respect to the axis of separation of the static quark-antiquark pair, i.e.\ with respect to the $z$ axis.

\item $\eta = +,-$, the eigenvalue corresponding to the operator $\mathcal{P} \circ \mathcal{C}$, i.e.\ the combination of parity and charge conjugation.

\item $\epsilon = +,-$, the eigenvalue corresponding to the operator $\mathcal{P}_x$, which denotes the spatial reflection along the $x$ axis, which is perpendicular to the axis of separation of the static quark-antiquark pair.
\end{itemize}
It is conventional to write $\Lambda = \Sigma,\Pi,\Delta$ instead of $\Lambda = 0,1,2$ and $\eta = g,u$ instead of $\eta = +,-$. Note that for angular momentum $\Lambda > 0$ the spectrum is degenerate with respect to $\epsilon = +$ and $\epsilon = -$. The labeling of states is thus $\Lambda^{\epsilon}_{\eta}$ for $\Lambda = 0 = \Sigma$ and $\Lambda_{\eta}$ for $\Lambda > 0$. For a more detailed discussion of those quantum numbers cf.\ e.g.\ \cite{Bali:2005fu,Bicudo:2015kna}.


\subsection{Angular momentum $\Lambda$}\label{ssec:1-2}

We start in the continuum and consider hybrid static potential creation operators and trial states
\begin{eqnarray}
\label{EQN895} \underbrace{\ket{\Psi_\text{hybrid}}_{S;\Lambda}}_{\textrm{trial state}} \ \ = \ \ \underbrace{\int_0^{2\pi} d\varphi \, \textrm{exp}(i \Lambda \varphi) R
(\varphi) O_S}_{\textrm{creation operator}} \ket{\Omega} ,
\end{eqnarray}
where $\ket{\Omega}$ is the vacuum and $R(\varphi)$ denotes a rotation by an angle $\varphi$ around the $z$ axis. Moreover,
\begin{eqnarray}
\label{eq:psiS} O_S \ket{\Omega} \ \ = \ \ \bar{Q}(-r/2) U(-r/2,r_1) S(r_1,r_2) U(r_2,+r/2) Q(+r/2) \ket{\Omega} ,
\end{eqnarray}
where $Q(+r/2)$ and $\bar{Q}(-r/2)$ are operators creating a spinless quark-antiquark pair and \\ $U(-r/2,r_1) S(r_1,r_2) U(r_2,+r/2)$ is a parallel transporter connecting the quark and the antiquark in a gauge invariant way. $U(-r/2,r_1)$ and $U(r_2,+r/2)$ denote straight parallel transporters along the $z$ axis (in the simplest case $r_1 = -r/2$ and $r_2 = +r/2$, i.e.\ $U(-r/2,r_1) = U(r_2,+r/2) = 1$), while the operator $S(r_1,r_2)$ is different from a straight line and, thus, generates a gluonic excitation. It is easy to show that the trial state (\ref{EQN895}) has definite angular momentum $\Lambda$ (see appendix~\ref{APP123}).

The corresponding lattice expression is
\begin{eqnarray}
\label{eq:L} \ket{\Psi_\text{hybrid}}_{S;\Lambda} \ \ = \ \ \sum_{k = 0}^3 \textrm{exp}\bigg(\frac{i \pi \Lambda k}{2}\bigg) R\bigg(\frac{\pi k}{2}\bigg) O_S \ket{\Omega} ,
\end{eqnarray}
where the angle of rotation is restricted to multiples of $\pi/2$ and $U(-r/2,r_1)$, $S(-r/2,+r/2)$ and $U(r_2,+r/2)$ are products of gauge links. For example for $\Lambda = 1$
\begin{eqnarray}
\ket{\Psi_\text{hybrid}}_{S;\Lambda=1} \ \ = \ \ \bigg(1 + i R\bigg(\frac{\pi}{2}\bigg) - R(\pi) - i R\bigg(\frac{3\pi}{2}\bigg)\bigg) O_s \ket{\Omega} ,
\end{eqnarray}
i.e.\ one has to compute Wilson loops, where each of the straight spatial Wilson lines is replaced by a sum over the four rotations of the operators $O$ with weight factors $+1$, $+i$, $-1$ and $-i$.

Note that, due to the restriction to cubic rotations, the lattice trial states do not have definite angular momentum. They receive contributions from an infinite number of angular momentum sectors as follows (for details cf.\ standard textbooks on group theory, e.g.\ \cite{Cornwell:1997ke}):
\begin{itemize}
\item $\Lambda = \Sigma$ corresponds to absolute angular momenta $\{ 0,4,8,12,\ldots \}$,

\item $\Lambda = \Pi$ corresponds to absolute angular momenta $\{ 1,3,5,7,\ldots \}$,

\item $\Lambda = \Delta$ corresponds to angular momenta $\{ 2,6,10,14,\ldots \}$.
\end{itemize}


\subsection{$\mathcal{P} \circ \mathcal{C}$ and $\mathcal{P}_x$ quantum numbers $\eta$ and $\epsilon$}

It is straightforward to show
\begin{eqnarray}
\nonumber & & \hspace{-0.7cm} (\mathcal{P} \circ \mathcal{C}) O_S \ket{\Omega} \ \ = \ \ (\mathcal{P} \circ \mathcal{C}) \bar{Q}(-r/2) U(-r/2,r_1) S(r_1,r_2) U(r_2,+r/2) Q(+r/2) \ket{\Omega} \ \ = \\
 & & = \ \ \bar{Q}(-r/2) U(-r/2,-r_2) S_{\mathcal{P} \circ \mathcal{C}}(-r_2,-r_1) U(-r_1,+r/2) Q(+r/2) \ket{\Omega} ,
\end{eqnarray}
where $S_{\mathcal{P} \circ \mathcal{C}}(-r_2,-r_1)$ is the charge conjugated spatial reflection of $S(r_1,r_2)$ with respect to the center of the separation axis. Consequently, one has to include both $S$ and $S_{\mathcal{P} \circ \mathcal{C}}$ in the final operator, to obtain a trial state with definite $\eta$. Similarly,
\begin{eqnarray}
\nonumber & & \hspace{-0.7cm} \mathcal{P}_x O_S \ket{\Omega} \ \ = \ \ \mathcal{P}_x \bar{Q}(-r/2) U(-r/2,r_1) S(r_1,r_2) U(r_2,+r/2) Q(+r/2) \ket{\Omega} \ \ = \\
 & & = \ \ \bar{Q}(-r/2) U(-r/2,r_1) S_{\mathcal{P}_x}(r_1,r_2) U(r_2,+r/2) Q(+r/2) \ket{\Omega} ,
\end{eqnarray}
where $S_{\mathcal{P}_x}(r_1,r_2)$ is the spatial reflection of $S(r_1,r_2)$ along the $x$ axis.

\begin{figure}[htb]
\begin{center}
\includegraphics{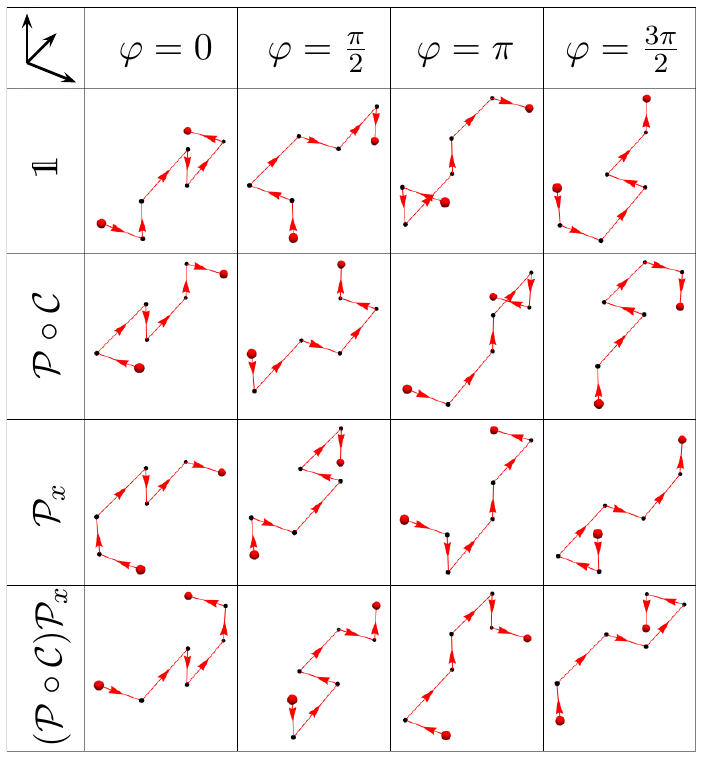}
\end{center}
\caption{\label{fig:trafotable}Terms appearing in the construction of the trial state via eq.\ (\ref{eq:trialstate}) for an exemplary operator $S(-r/2,+r/2)$ (top left). The columns correspond to rotations of the operator around the separation axis, while the rows correspond to applications of $\mathcal{P} \circ \mathcal{C}$ and $\mathcal{P}_x$. Red lines represent gauge link variables, red spheres the quark and the antiquark and black dots lattice sites.}
\end{figure}

To construct a trial state, which has definite quantum numbers $\Lambda_\eta^\epsilon$, we take the state (\ref{eq:L}), which has angular momentum $\Lambda$, and project that state onto the subspace of eigenstates of the operators $\mathcal{P} \circ \mathcal{C}$ and $\mathcal{P}_x$ characterized by $\eta$ and $\epsilon$, respectively:
\begin{eqnarray}
\nonumber & & \hspace{-0.7cm} \ket{\Psi_\text{hybrid}}_{S;\Lambda_\eta^\epsilon} \ \ = \ \ \mathds{P}_{P C,\eta} \mathds{P}_{P_x,\epsilon} \ket{\Psi_\text{hybrid}}_{S;\Lambda} \ \ = \\
\nonumber & & = \ \ \frac{1}{4} \Big(1 + \eta (\mathcal{P} \circ \mathcal{C}) + \epsilon \mathcal{P}_x + \eta \epsilon (\mathcal{P} \circ \mathcal{C}) \mathcal{P}_x\Big) \sum_{k=0}^3 \textrm{exp}\bigg(\frac{i \pi \Lambda k}{2}\bigg) R\bigg(\frac{\pi k}{2}\bigg) O_S \ket{\Omega} \ \ = \\
\label{eq:trialstate} & & = \ \ \bar{Q}(-r/2) a_{S;\Lambda_\eta^\epsilon}(-r/2,+r/2) Q(+r/2) \ket{\Omega}
\end{eqnarray}
with projectors
\begin{eqnarray}
\mathds{P}_{P C,\eta} \ \ = \ \ \frac{1}{2}(1 + \eta (\mathcal{P} \circ \mathcal{C})) \quad , \quad \mathds{P}_{P_x,\epsilon} \ \ = \ \ \frac{1}{2} (1 + \epsilon \mathcal{P}_x)
\end{eqnarray}
and
\begin{eqnarray}
\nonumber & & \hspace{-0.7cm} a_{S;\Lambda_\eta^\epsilon}(-r/2,+r/2) \ \ = \\
\nonumber & & = \ \ \frac{1}{4} \sum_{k=0}^3 \textrm{exp}\bigg(\frac{i \pi \Lambda k}{2}\bigg) R\bigg(\frac{\pi k}{2}\bigg)
\Big(U(-r/2,r_1) \Big(S(r_1,r_2) + \epsilon S_{\mathcal{P}_x}(r_1,r_2)\Big) U(r_2,+r/2) + \\
\label{EQN644} & & \hspace{0.675cm} U(-r/2,-r_2) \Big(\eta S_{\mathcal{P} \circ \mathcal{C}}(-r_2,-r_1) + \eta \epsilon S_{(\mathcal{P} \circ \mathcal{C}) \mathcal{P}_x}(-r_2,-r_1)\Big) U(-r_1,+r/2)\Big) .
\end{eqnarray}
Notice that not every operator $S(r_1,r_2)$ is suited to construct trial states for any given set of quantum numbers $\Lambda_\eta^\epsilon = \Lambda'^{\epsilon'}_{\eta'}$, i.e.\ for some $\Lambda'^{\epsilon'}_{\eta'}$ the trial state defined in (\ref{eq:trialstate}) is zero, $\ket{\Psi_\text{hybrid}}_{S;\Lambda'^{\epsilon'}_{\eta'}} = 0$. Also note, that even though sectors with $\Lambda \geq 1$ (i.e.\ in this work the $\Pi$ and the $\Delta$ sectors) are degenerate with respect to the quantum number $\epsilon$, creation operators constructed via eq.\ (\ref{eq:trialstate}) with either $\epsilon = +$ or $\epsilon = -$ might be different, i.e.\ yield non-identical correlation functions. In such cases it is important to consider the $\epsilon = +$ and $\epsilon = -$ operators as separate creation operators, when identifying an optimal set of creation operators for each correlation matrix (cf.\ section~\ref{SEC450}; an example is the $\Pi_g$ sector, when using $S_{IV,1}$). In practice, we use eq.\ (\ref{eq:trialstate}) to automatically generate creation operators with definite quantum numbers $\Lambda_\eta^\epsilon$ from all considered operators $S(r_1,r_2)$ (cf.\ Figure~\ref{fig:trafotable} for a graphical illustration of an example).

\subsection{Correlation functions}

We determine hybrid static potentials with quantum numbers $\Lambda_\eta^\epsilon$, which we denote by $V_{\Lambda_\eta^\epsilon}(r)$, from the asymptotic exponential behavior of temporal correlation functions
\begin{eqnarray}
\label{EQN599} W_{S,S';\Lambda_\eta^\epsilon}(r,t) \ \ = \ \ \bra{\Psi_\text{hybrid}(t)}_{S;\Lambda_\eta^\epsilon} \ket{\Psi_\text{hybrid}(0)}_{S';\Lambda_\eta^\epsilon} \ \ \sim_{t \rightarrow \infty} \ \ \exp\Big(-V_{\Lambda_\eta^\epsilon}(r) t\Big) .
\end{eqnarray}
Expressing eq.\ (\ref{EQN599}) in terms of a path integral and performing the integration over the static quarks leads to
\begin{eqnarray}
\nonumber & & \hspace{-0.7cm} W_{S,S';\Lambda_\eta^\epsilon}(r,t) \ \ = \\
\label{EQN600} & & = \ \ \bigg\langle \textrm{Tr}\Big(
  a_{S';\Lambda_\eta^\epsilon}(-r/2,+r/2;0)
  U(+r/2;0,t)
  \Big(a_{S;\Lambda_\eta^\epsilon}(-r/2,+r/2;t)\Big)^\dagger
  U(-r/2;t,0)
\Big) \bigg\rangle_U ,
\end{eqnarray}
where $U(r;t_1,t_2)$ denotes a straight line of temporal gauge links at $r$ from time $t_1$ to $t_2$ and $\langle \ldots \rangle_U$ is the average on an ensemble of gauge link configurations distributed according to $e^{-S}$. The right hand side of this equation can be computed using standard techniques from lattice field theory as briefly summarized in section~\ref{SEC567}.


\newpage

\section{\label{SEC567}Lattice setup}

All computations presented in this work have been performed using SU(3) lattice gauge theory. The gauge link configurations have been generated with the standard Wilson gauge action (cf.\ standard textbooks on lattice field theory, e.g.\ \cite{Rothe:1992nt}) and the Chroma QCD library \cite{Edwards:2004sx}. Since we are considering purely gluonic observables, we expect that there is little difference between our SU(3) Yang-Mills results and corresponding results in full QCD (cf.\ also the discussion of systematic errors at the end of section~\ref{SEC755} and \cite{Bali:2000vr}).

In this work we use a single ensemble with lattice extent $24^3 \times 48$ and gauge coupling $\beta = 6.0$ corresponding to lattice spacing $a \approx 0.093 \, \textrm{fm}$ and spacetime volume $ \approx (2.22 \, \textrm{fm})^3 \times 4.44 \, \textrm{fm}$, when identifying $r_0$ with $0.5 \, \textrm{fm}$ (for a determination of $r_0$ cf.\ section~\ref{SEC679}). The gauge link configurations are separated by 20 lattice updates, where each update comprises a heatbath and four over-relaxation steps. We have performed standard binning analyses with bins containing either 1, 2 or 4 gauge link configurations. We have found that the statistical errors of the $\Sigma_g^+$ potential are essentially independent of the bin size, which indicates that performing 20 lattice updates largely eliminates correlations in Monte Carlo time. For the final results for hybrid static potentials presented in section~\ref{SEC499} and Table~\ref{TAB400} we have generated more than $5 \, 500$ gauge link configurations. During the time-consuming optimization of hybrid static potential creation operators and trial states discussed in section~\ref{SEC445}, we use a subset of 100 gauge link configurations, to reduce the computational effort to an acceptable level.

To improve the signal quality, standard smearing techniques are applied to the gauge links of the Wilson loop-like correlation functions (\ref{EQN599}). The temporal gauge links in $U(r;t_1,t_2)$ are HYP2 smeared gauge links \cite{Hasenfratz:2001hp,DellaMorte:2003mn,Della Morte:2005yc}, which lead to a reduced self energy of the static quarks and, consequently, to smaller statistical errors. The spatial gauge links in $a_{S;\Lambda_\eta^\epsilon}(r_1,r_2;t)$ are APE smeared gauge links (for detailed equations cf.\ e.g.\ \cite{Jansen:2008si}), where the parameters are tuned to optimize the ground state overlaps (cf.\ section~\ref{SEC802}) and, thus, allow to extract the potentials at smaller temporal separations.

All statistical errors shown and quoted throughout this paper, e.g.\ for the hybrid static potentials in section~\ref{SEC499} or the potential parameterizations and hybrid meson masses in section~\ref{SEC498}, are determined via an evolved jackknife analysis starting at the level of the correlation matrices. To exclude statistical correlations between gauge link configurations, which are close in Monte Carlo simulation time, we perform a suitable binning of these configurations.


\newpage

\section{\label{SEC445}Optimization of hybrid static potential creation operators and trial states}

Since the signal-to-noise ratio of correlation functions (\ref{EQN599}) decreases exponentially with respect to the temporal separation, it is essential to identify hybrid static potential creation operators, which generate trial states with large ground state overlap. This allows to extract hybrid static potentials at rather small temporal separations, where the signal-to-noise ratio is favorable.

\begin{figure}[p]
\begin{center}
\includegraphics[scale=0.92]{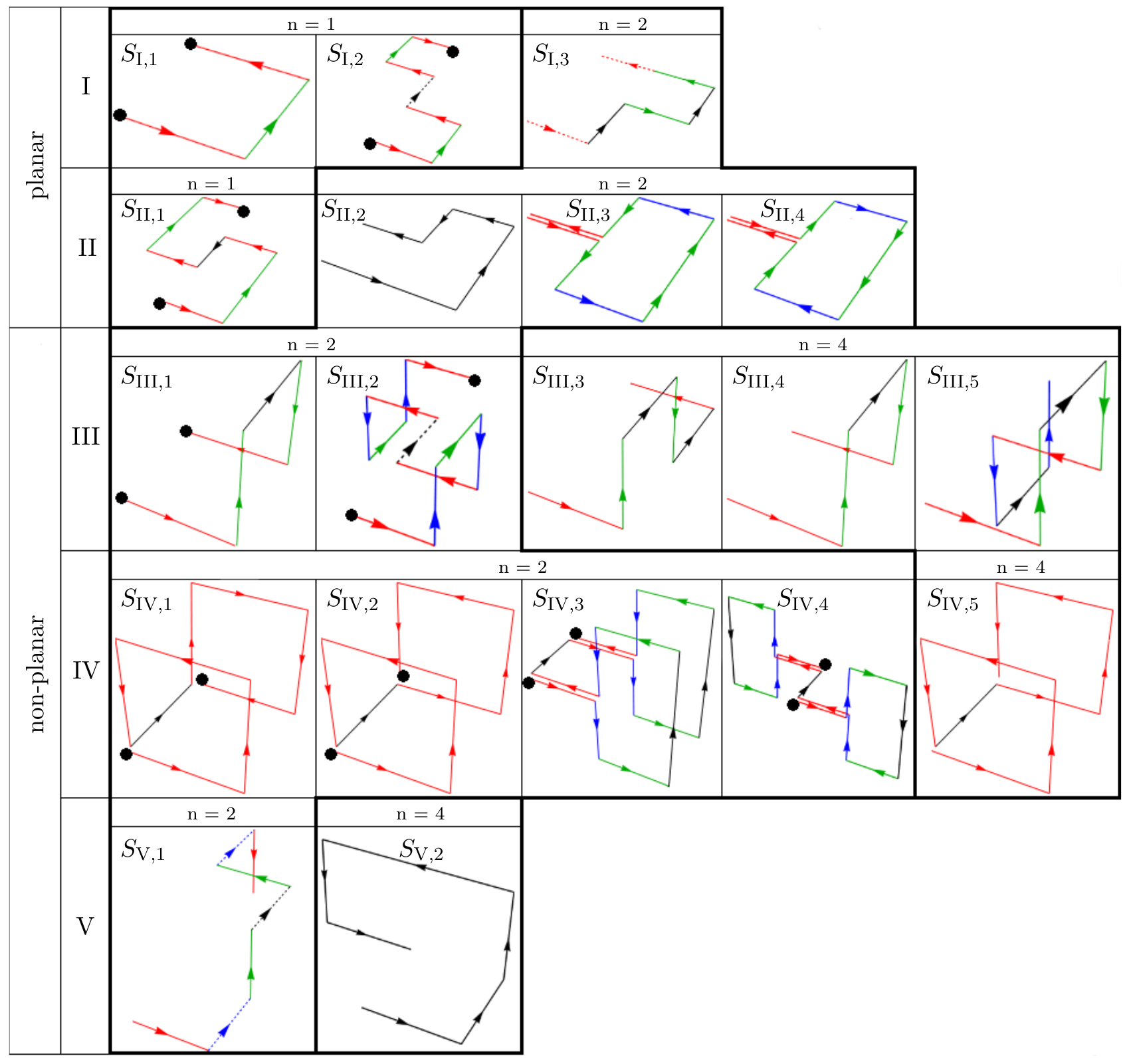}
\end{center}
\caption{\label{fig:0shapes}Operators $S$ used to generate trial states $\ket{\Psi_\text{hybrid}}_{S;\Lambda_\eta^\epsilon}$ according to eqs.\ (\ref{eq:trialstate}) to (\ref{EQN644}). Each arrow represents a straight path of gauge links. Arrows with the same color (red, green or blue) have the same length, i.e.\ represent the same number of gauge links. Dotted arrows can have length zero, while solid arrows represent at least one gauge link. If the starting point and the end point of $S$ are marked by black dots, $U(-r/2,r_1)$ and $U(r_2,+r/2)$ in eq.\ (\ref{eq:psiS}) have the same length, i.e.\ $r_1 - (-r/2) = +r/2 - r_2$, else their length can be different. $n \in \{1,2,4\}$ is the number of differently oriented operators (i.e.\ operators, which cannot be transformed into each other by rotations around the $z$ axis) obtained by applying $\mathcal{P}\circ \mathcal{C}$, $\mathcal{P}_x$ and $(\mathcal{P}\circ \mathcal{C}) \mathcal{P}_x$.}
\end{figure}

The starting point is a large set of quite distinct operators $S$, some of them simple, others of more complicated shape, which are shown in Figure~\ref{fig:0shapes}. All these operators extend over regions, which are of the same order as the quark-antiquark separation. This is quite different from our previous exploratory study in SU(2) Yang-Mills theory \cite{Wolf:2014tta}, where we used local chromoelectric and chromomagnetic field strength insertions. While the latter are theoretically easier to handle and thus are quite common in analytical studies, e.g.\ based on pNRQCD \cite{Berwein:2015vca}, the extended operators we are using here are much better suited for numerical lattice field theory studies, because they lead to trial states with larger ground state overlaps and, thus, to results with significantly smaller statistical errors.

The operators $S$ can be categorized into planar operators,
\begin{itemize}
\item[(I)] where gauge links parallel to the $z$ axis are exclusively pointing in positive $z$ direction (as before the static quark-antiquark pair is separated along the $z$ axis),

\item[(II)] with gauge links parallel to the $z$ axis both in positive and negative $z$ direction,
\end{itemize}
and into non-planar operators,
\begin{itemize}
\item[(III)] without closed loops,

\item[(IV)] with closed loops,

\item[(V)] with spiral-like structures.
\end{itemize}
Some of these operators, e.g.\ $S_{\text{I},1}$, $S_{\text{I},3}$, $S_{\text{III},1}$, $S_{\text{III},2}$, $S_{\text{III},4}$, $S_{\text{III},5}$, $S_{\text{IV},1}$, $S_{\text{IV},2}$, $S_{\text{IV},3}$, $S_{\text{IV},5}$ and $S_{\text{V},2}$ were already used in previous lattice field theory studies of hybrid mesons, e.g.\ in \cite{Juge:1997nc}, while other operators are explored for the first time in this work. From these operators we construct a large number of different trial states $\ket{\Psi_\text{hybrid}}_{S;\Lambda_\eta^\epsilon}$ using eqs.\ (\ref{eq:trialstate}) to (\ref{EQN644}).

To check, whether a trial state $\ket{\Psi_\text{hybrid}}_{S;\Lambda_\eta^\epsilon}$ has large ground state overlap, we compute the effective mass
\begin{eqnarray}
\label{EQN632} V_{\textrm{eff};S;\Lambda_\eta^\epsilon}(r,t) a \ \ = \ \ \ln\bigg(\frac{W_{S,S;\Lambda_\eta^\epsilon}(r,t)}{W_{S,S;\Lambda_\eta^\epsilon}(r,t+a)}\bigg)
\end{eqnarray}
at small temporal separations, in particular at $t = a$, where contributions of excited states are most prominent. Small effective masses indicate trial states with large ground state overlaps, while operators leading to large effective masses can be discarded. In the following we discuss in detail, how we identify and optimize a small set of relevant operators $S$ for each hybrid static potential sector $\Lambda_\eta^\epsilon$.


\subsection{\label{SEC477}Optimization of the extents of the operators $S$}

In a first step we consider each of the operators $S$ shown in Figure~\ref{fig:0shapes} separately and optimize their extents for each hybrid potential sector $\Lambda_\eta^\epsilon$. In other words, for each arrow in Figure~\ref{fig:0shapes} we determine the number of gauge links it represents, such that the ground state overlap of the corresponding trial state is maximal.

\begin{figure}[htb]
\begin{center}
\includegraphics[scale=0.82]{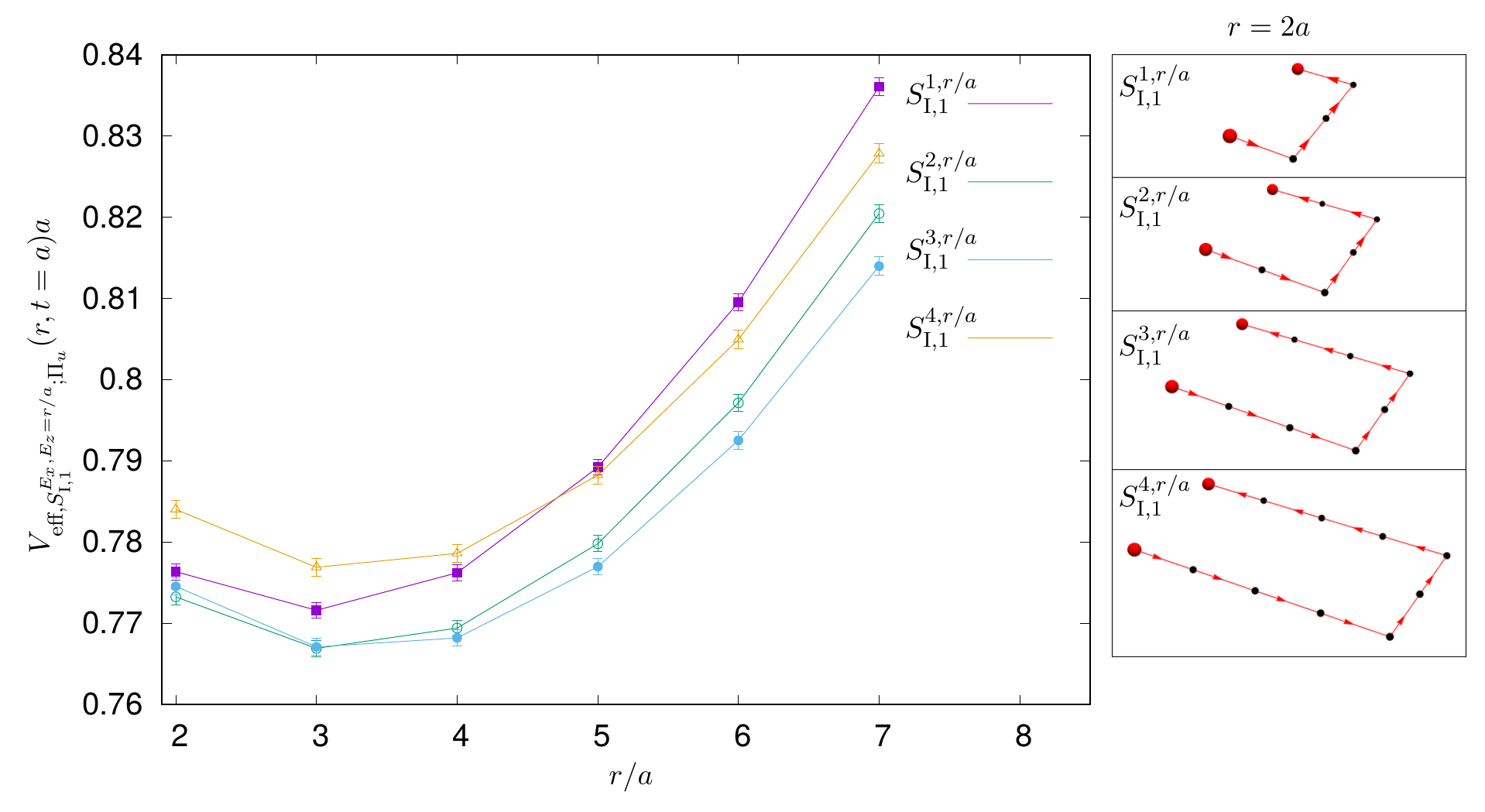}
\end{center}
\caption{\label{fig:S1_xy}Investigation of the dependence of $V_{\textrm{eff};S_{\text{I},1}^{E_x,E_z = r/a};\Pi_u}(r,t=a)$ on $E_x$. Red spheres, red arrows, and black dots represent quarks, gauge links and lattice sites, respectively.}
\end{figure}

\begin{figure}[htb]
\begin{center}
\includegraphics{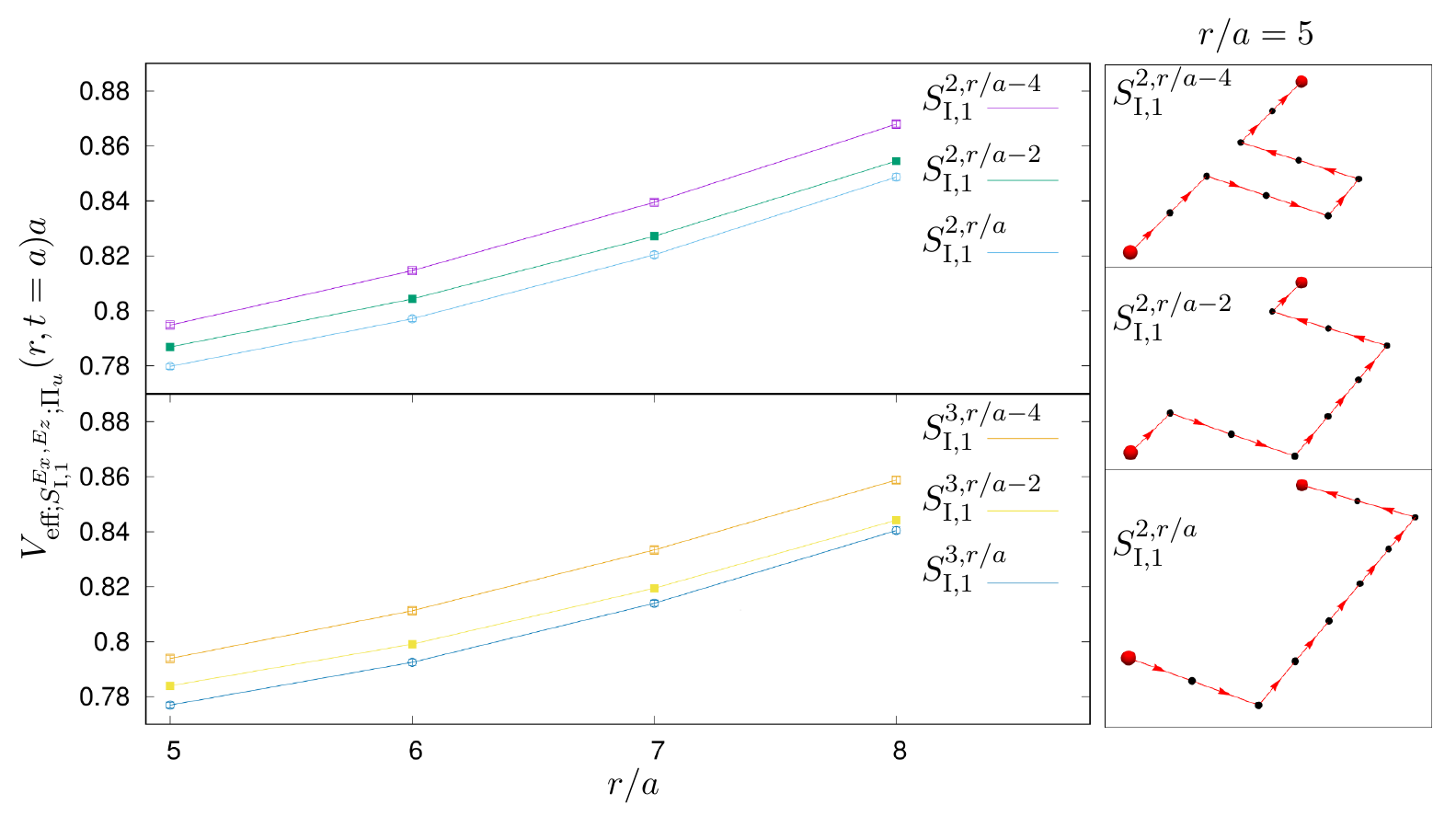}
\end{center}
\caption{\label{fig:S1_z}Investigation of the dependence of $V_{\textrm{eff};S_{\text{I},1}^{E_x,E_z};\Pi_u}(r,t=a)$ on $E_z$ ($E_x \in \{ 2,3 \}$, the optimum according to Figure~\ref{fig:S1_xy}). Red spheres, red arrows, and black dots represent quarks, single gauge links and lattice sites, respectively.}
\end{figure}

As an example we briefly discuss the optimization of the operator $S_{\text{I},1}$ for the $\Pi_u$ hybrid static potential. Variations of $S_{\text{I},1}$ are denoted by $S_{\text{I},1}^{E_x,E_z}$, where $E_x$ and $E_z$ are the operator extents in units of the lattice spacing in the $x$ direction and the $z$ direction, respectively. To keep the computational cost of the optimization on a feasible level, we first determine the optimal value for $E_x$ and after that the optimal value for $E_z$. Figure~\ref{fig:S1_xy} shows that the optimal value for $E_x$ weakly depends on the separation of the quark-antiquark pair $r$. For $r/a \leq 3$ the operator extent $E_x = 2$ minimizes the effective mass $V_{\textrm{eff};S_{\text{I},1}^{E_x,E_z};\Pi_u}(r,t=a)$ and, thus, the corresponding trial state has better overlap to the ground state in the $\Pi_u$ sector than trial states generated with $E_x \in \{ 1,2,4 \}$. Similarly, for $r/a \geq 4$ the operator with extent $E_x = 3$ minimizes $V_{\textrm{eff};S_{\text{I},1}^{E_x,E_z};\Pi_u}(r,t=a)$ and, thus, maximizes the ground state overlap. Since operator extents $E_x \in \{ 1,4 \}$ do not minimize $V_{\textrm{eff};S_{\text{I},1}^{E_x,E_z};\Pi_u}(r,t=a)$ for any of the considered quark-antiquark separations $r$, they are discarded. In Figure~\ref{fig:S1_z} we show an analogous comparison of effective masses $V_{\textrm{eff};S_{\text{I},1}^{E_x,E_z};\Pi_u}(r,t=a)$ for different $E_z$ and the previously optimized $E_x \in \{ 2,3\}$. We find that the optimum is $E_z = r/a$ independent of $r$ and $E_x$, i.e.\ the $z$ extent of operator $S_{\text{I},1}$ should be identical to the quark-antiquark separation, when used to compute the ground state hybrid static potential in the $\Pi_u$ sector.

All 19 operators $S$ shown in Figure~\ref{fig:0shapes} are optimized for each sector $\Lambda_\eta^\epsilon$ and each separation $r$ in a similar way. In the majority of cases more than two extents have to be optimized.


\subsection{\label{SEC802}Optimization of APE smearing parameters}

\begin{figure}[htb]
\begin{center}
\includegraphics{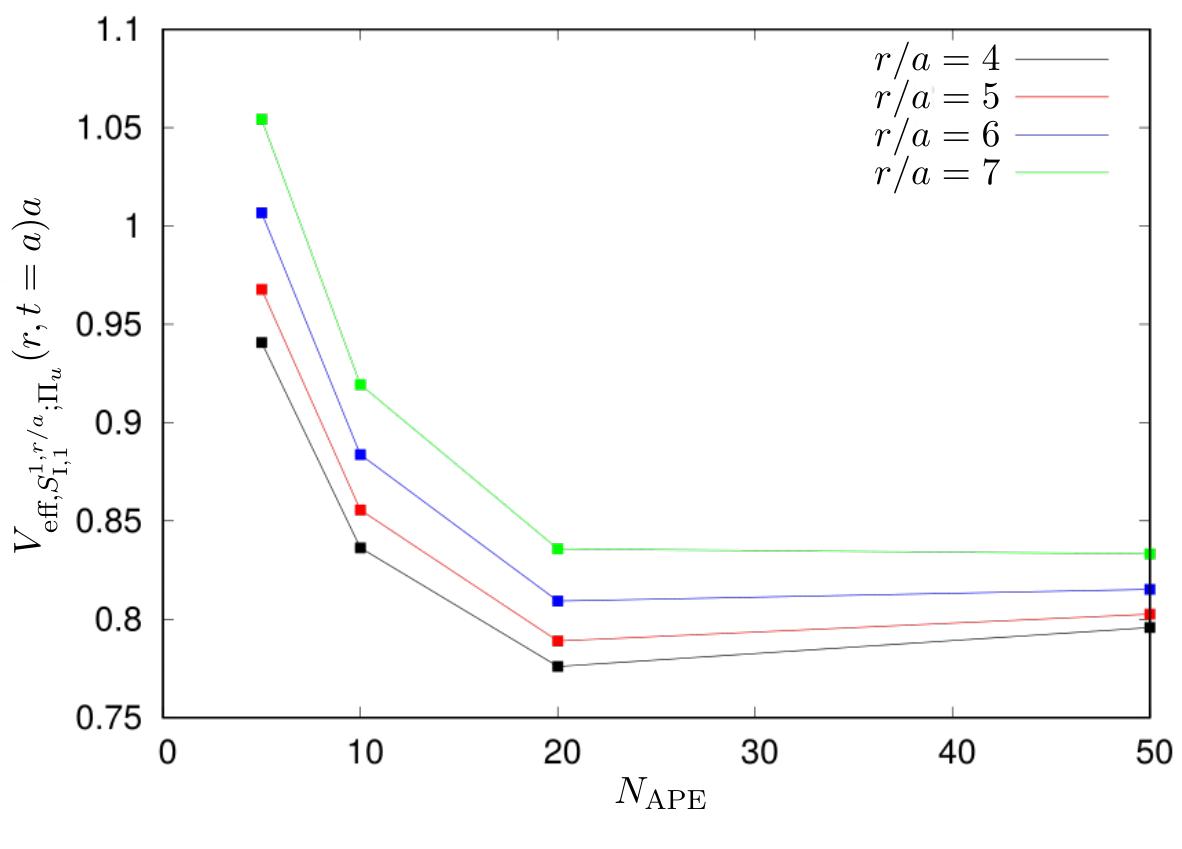}
\end{center}
\caption{\label{fig:APEinit}Investigation of the dependence of $V_{\textrm{eff};S_{\text{I},1}^{1,r/a};\Pi_u}(r,t=a)$ on the number of APE smearing steps $N_\textrm{APE}$.}
\end{figure}

To further improve the ground state overlap of the trial states, we use APE smeared spatial gauge links in $a_{S;\Lambda_\eta^\epsilon}(r_1,r_2;t)$ in (\ref{EQN600}). For detailed equations cf.\ e.g.\ \cite{Jansen:2008si}. We set $\alpha_\textrm{APE} = 0.5$, which is a common choice in the literature and we investigate the dependence of $V_{\textrm{eff};S;\Lambda_\eta^\epsilon}(r,t=a)$ on the number of APE smearing steps $N_\textrm{APE}$. An example plot for operator $S_{I,1}^{1,r/a}$ and the $\Pi_u$ hybrid static potential is shown in Figure~\ref{fig:APEinit}. While there is a significant increase of the ground state overlap, when increasing $N_\textrm{APE}$ from $0$ to around $20$, there is no further gain, when using $N_\textrm{APE} > 20$. This behavior is observed for various quark antiquark separations $r = 2 a, \ldots , 8 a$. Similar findings are obtained also for the other operators $S$ shown in Figure~\ref{fig:0shapes} and for all sectors $\Lambda_\eta^\epsilon$. Therefore, we use $N_\text{APE} = 20$ for all computations presented throughout this paper.

\subsection{\label{SEC450}Selecting optimal sets of trial states}

To further improve the ground state overlaps of the trial states, we resort to variational techniques for our final analyses in section~\ref{SEC499}. For each sector $\Lambda_\eta^\epsilon$ we use an ``optimal set'' of operators $S$, compute the corresponding correlation matrix (\ref{EQN600}) and solve generalized eigenvalue problems (see e.g.\ \cite{Luscher:1990ck,Blossier:2009kd}). In this way the static potentials are determined using an optimized linear combination of creation operators. To keep the computational effort on an acceptable level, we have restricted these variational analyses to the three or four most promising operators $S$ for each sector $\Lambda_\eta^\epsilon$ and separation $r$.

To select these operators, we have first performed an optimization of the extents of each operator as discussed in section~\ref{SEC477}. We have then taken those three or four operators, which yield the smallest effective masses (\ref{EQN632}) at $t = a$. Results are collected in the Table~\ref{TAB_Sgm} to Table~\ref{TAB_Du}. Each table corresponds to another hybrid static potential $\Lambda_\eta^\epsilon$. The operators are sketched in the left column of each table. In two cases ($\Lambda_\eta^\epsilon = \Sigma_u^+$ and $\Lambda_\eta^\epsilon = \Pi_g$, i.e.\ Table~\ref{TAB_Sup} and Table~\ref{TAB_Pg}) not only the extents, but also the operators $S$ change with the separation $r$, indicated by ``-''. The operators are also defined mathematically in the tables. For example the left hand side of $U S U = U_x^2 U_y^2 U_z^{E_z} U_{-y}^2 U_{-x}^2$ (first line of Table~\ref{TAB_Sgm}) represents $U(-r/2,r_1) S(r_1,r_2) U(r_2,+r/2)$ from eq.\ (\ref{EQN644}), while the right hand side denotes 2 links in positive $x$ direction, 2 links in positive $y$ direction, $E_z$ links in positive $z$ direction (where $E_z$ as a function of $r$ is listed directly below), 2 links in negative $y$ direction, 2 links in negative $x$ direction.


\begin{table}[p]
\begin{center}
\includegraphics{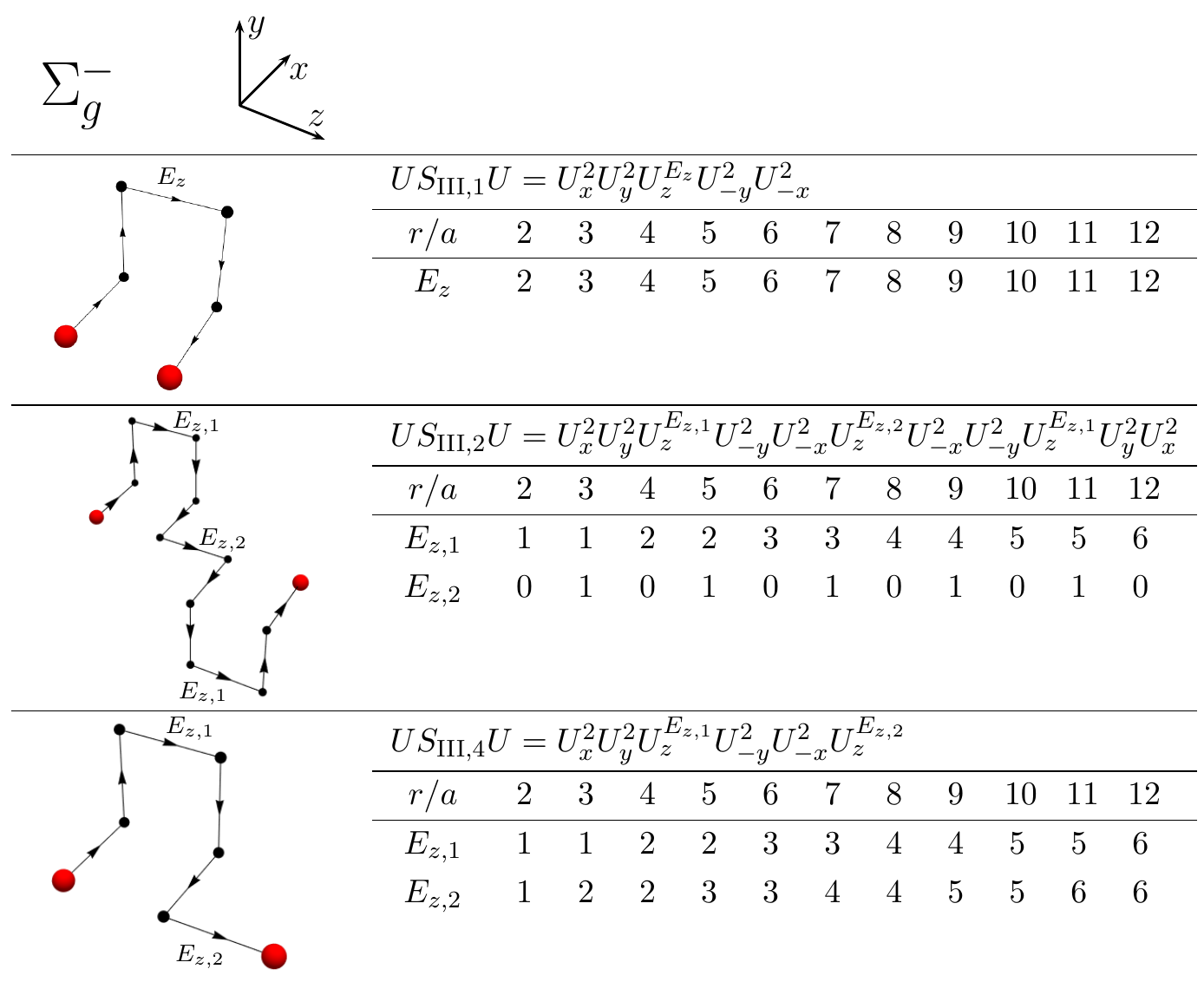}
\end{center}
\caption{\label{TAB_Sgm}Optimized creation operators for $V_{\Sigma_g^-}(r)$.}
\end{table}


\begin{table}[p]
\begin{center}
\includegraphics{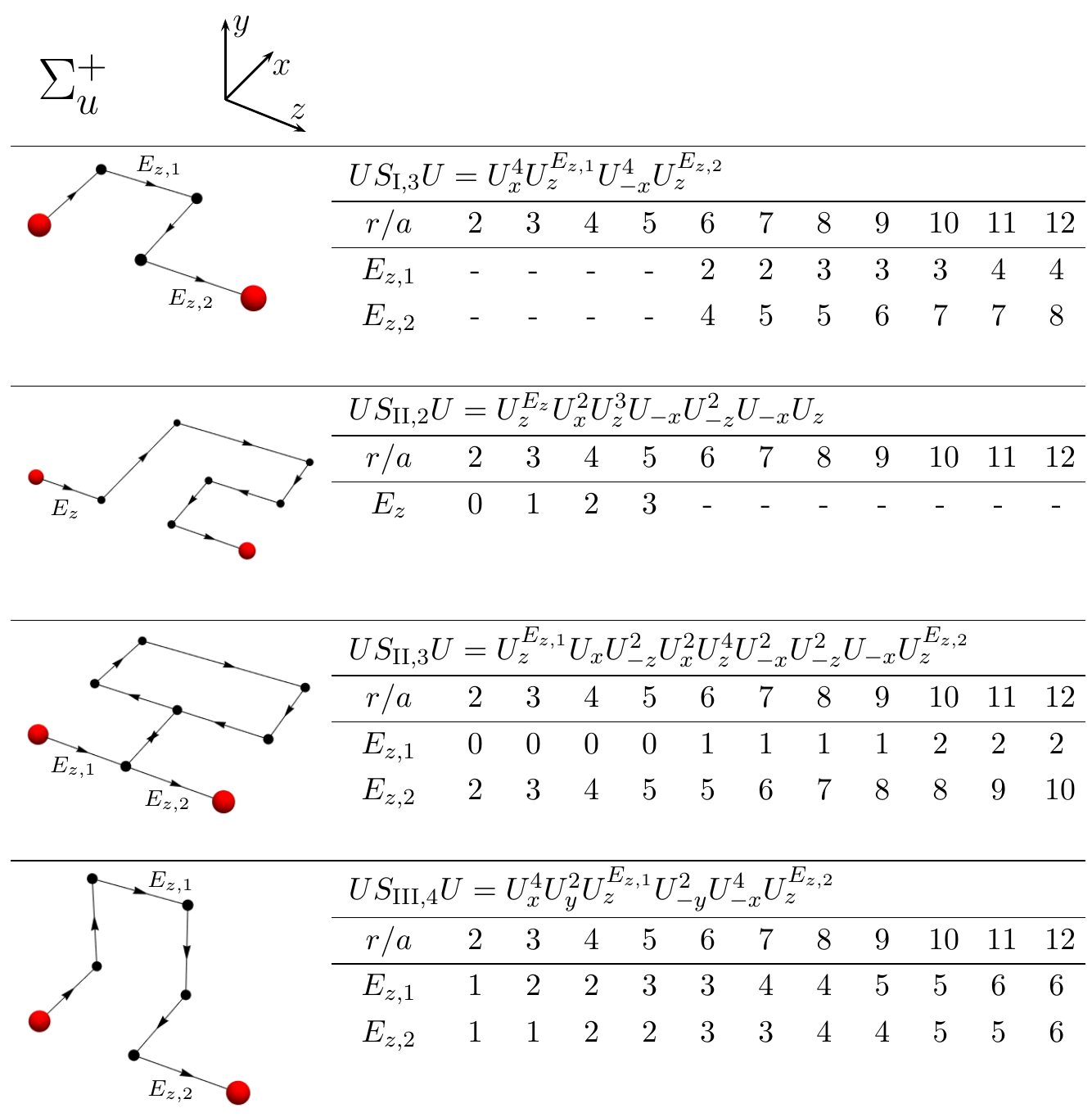}
\end{center}
\caption{\label{TAB_Sup}Optimized creation operators for $V_{\Sigma_u^+}(r)$.}
\end{table}


\begin{table}[p]
\begin{center}
\includegraphics{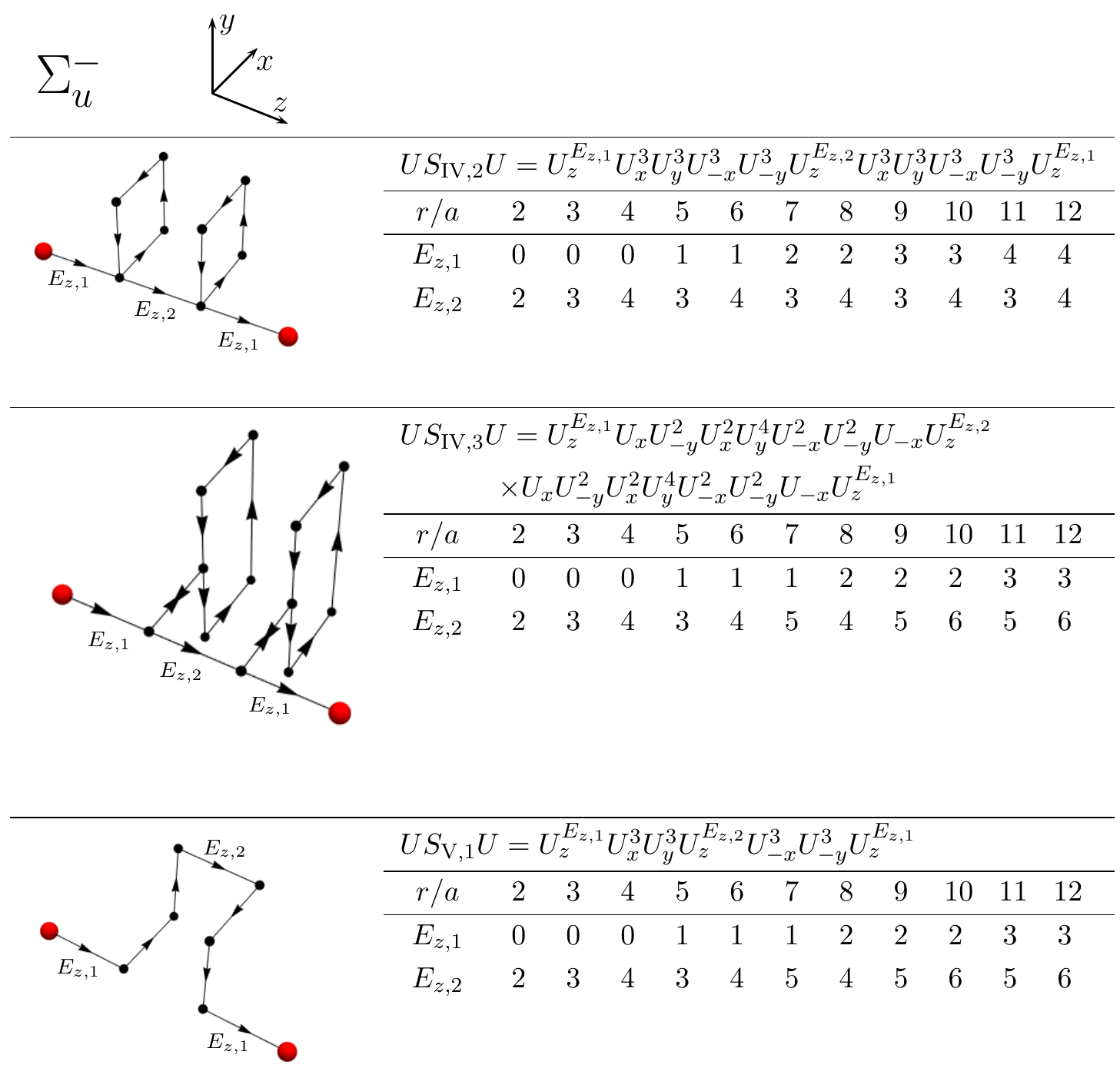}
\end{center}
\caption{\label{TAB_Sum}Optimized creation operators for $V_{\Sigma_u^-}(r)$.}
\end{table}


\begin{table}[p]
\begin{center}
\includegraphics{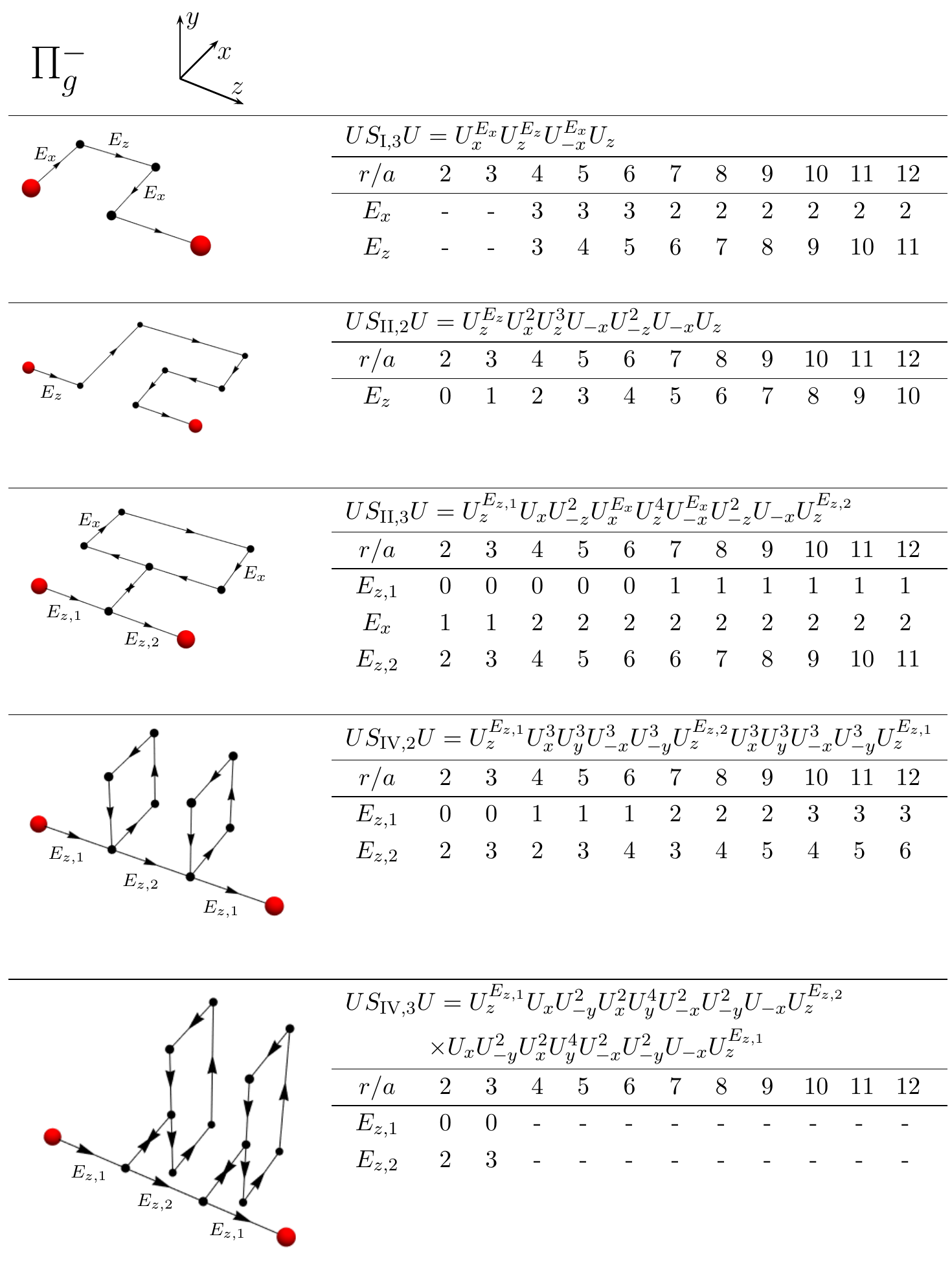}
\end{center}
\caption{\label{TAB_Pg}Optimized creation operators for $V_{\Pi_g}(r)$. Note that, even though the $\Pi_g$ hybrid potential is degenerate with respect to $\epsilon$, the construction of creation operators via eq.\ (\ref{eq:trialstate}) is not independent of $\epsilon$; the optimized set of creation operators corresponds to $\epsilon = -$ as indicated by $\Pi_g^-$ in the first line of the table.}
\end{table}


\begin{table}[p]
\begin{center}
\includegraphics{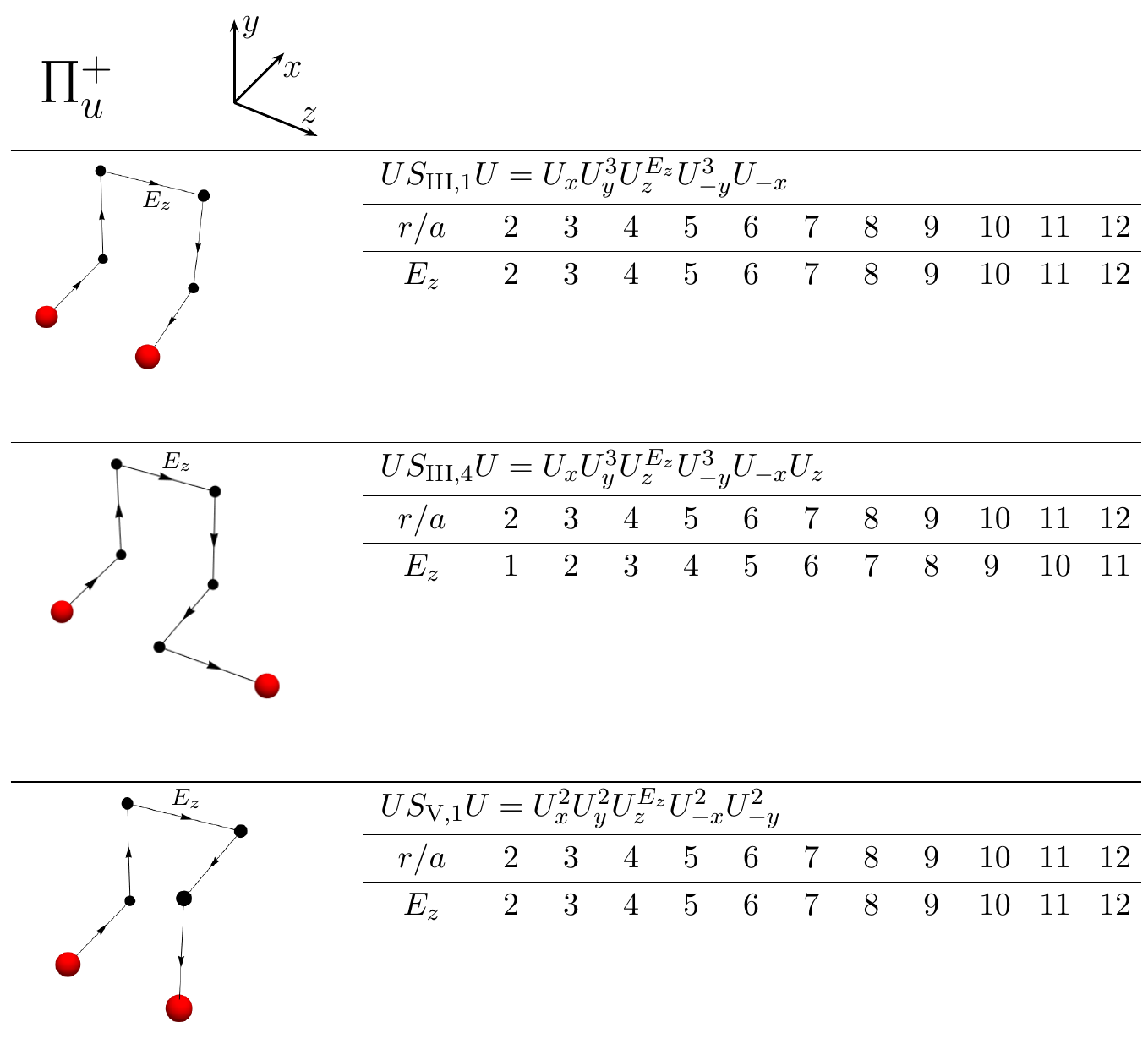}
\end{center}
\caption{\label{TAB_Pu}Optimized creation operators for $V_{\Pi_u}(r)$. Note that, even though the $\Pi_u$ hybrid potential is degenerate with respect to $\epsilon$, the construction of creation operators via eq.\ (\ref{eq:trialstate}) is not independent of $\epsilon$; the optimized set of creation operators corresponds to $\epsilon = +$ as indicated by $\Pi_u^+$ in the first line of the table.}
\end{table}


\begin{table}[p]
\begin{center}
\includegraphics{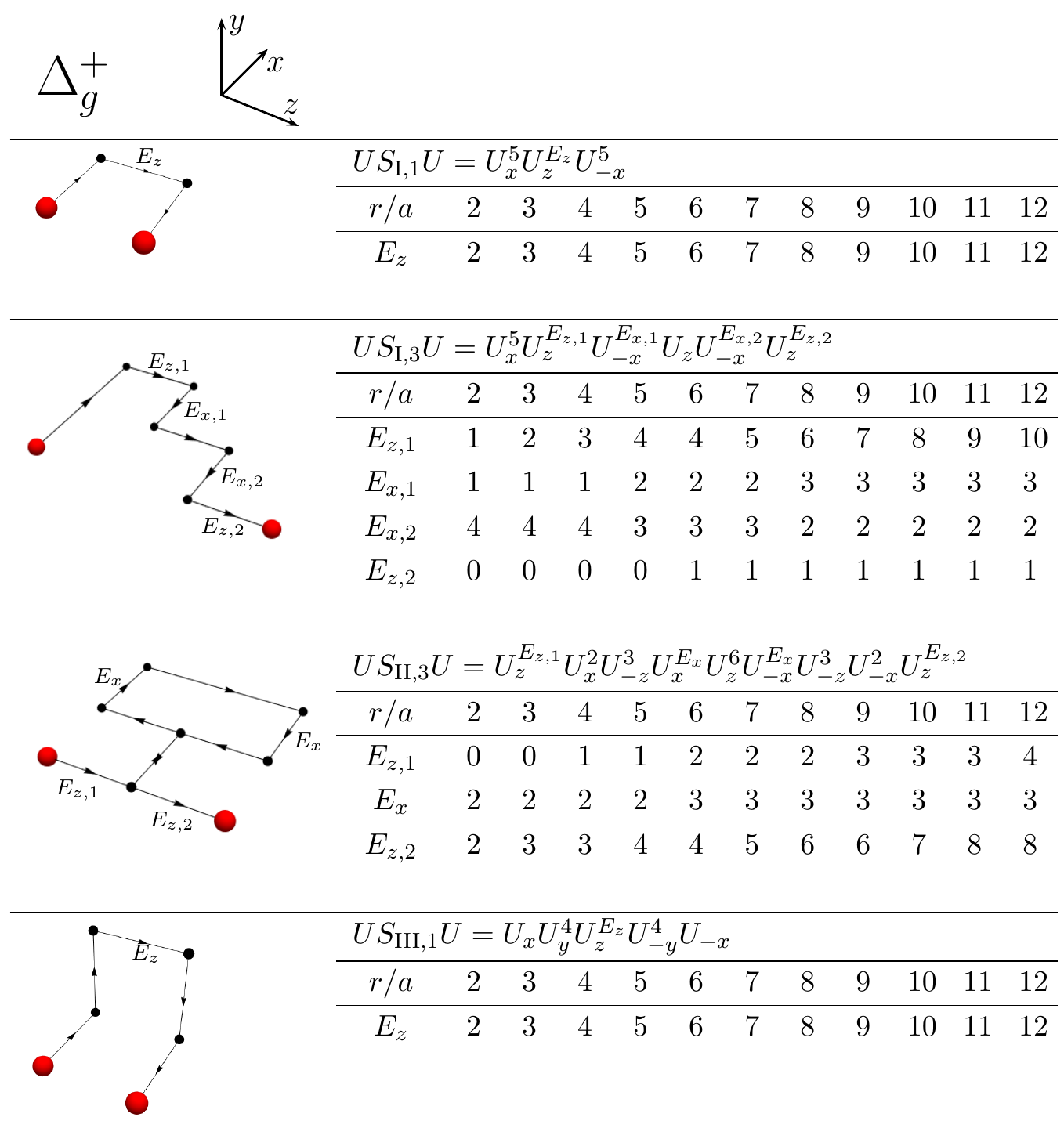}
\end{center}
\caption{\label{TAB_Dg}Optimized creation operators for $V_{\Delta_g}(r)$. Note that, even though the $\Delta_g$ hybrid potential is degenerate with respect to $\epsilon$, the construction of creation operators via eq.\ (\ref{eq:trialstate}) is not independent of $\epsilon$; the optimized set of creation operators corresponds to $\epsilon = +$ as indicated by $\Delta_g^+$ in the first line of the table.}
\end{table}


\begin{table}[p]
\begin{center}
\includegraphics{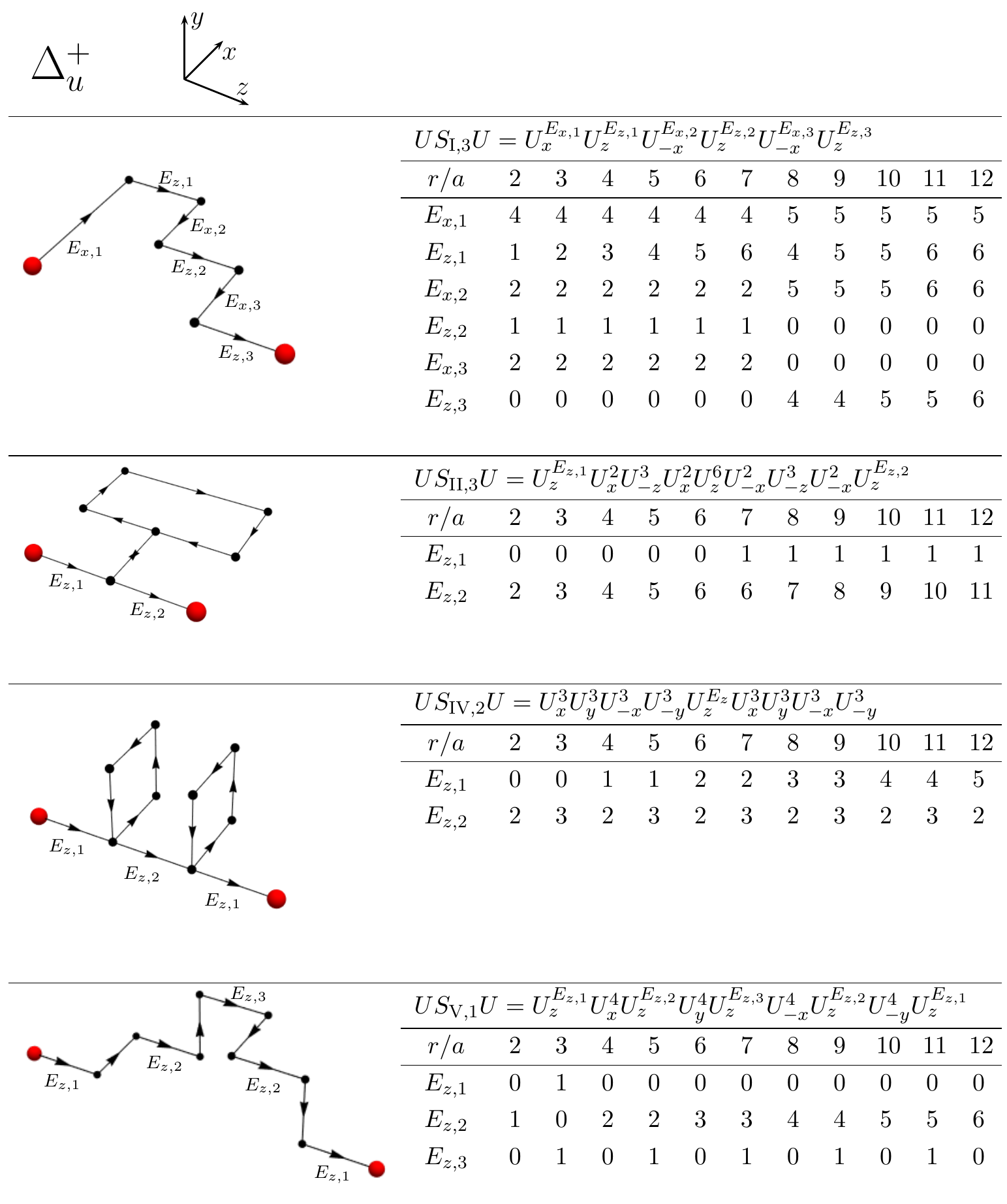}
\end{center}
\caption{\label{TAB_Du}Optimized creation operators for $V_{\Delta_u}(r)$. Note that, even though the $\Delta_u$ hybrid potential is degenerate with respect to $\epsilon$, the construction of creation operators via eq.\ (\ref{eq:trialstate}) is not independent of $\epsilon$; the optimized set of creation operators corresponds to $\epsilon = +$ as indicated by $\Delta_u^+$ in the first line of the table.}
\end{table}


\newpage

\section{\label{SEC499}Lattice field theory results for hybrid static potentials}

We compute the ground state hybrid static potential for each of the sectors \\ $\Lambda_\eta^\epsilon = \Sigma_g^-,\Sigma_u^+,\Sigma_u^-,\Pi_g,\Pi_u,\Delta_g,\Delta_u$ as well as the ground state and first excited static potential for the sector $\Lambda_\eta^\epsilon = \Sigma_g^+$. The latter is in the same energy region as the ground state hybrid static potentials and of particular interest, since it is expected to become degenerate with the $\Pi_g$ hybrid static potential in the limit of small quark-antiquark separations $r$. For these computations we use correlation matrices
\begin{eqnarray}
C_{j,k;\Lambda_\eta^\epsilon}(r,t) \ \ = \ \ W_{S_j,S_k;\Lambda_\eta^\epsilon}(r,t) .
\end{eqnarray}
\begin{itemize}
\item $\Sigma_g^-$, $\Sigma_u^+$, $\Sigma_u^-$, $\Pi_u$:
\\ $3 \times 3$ correlation matrices with operators as specified in Table~\ref{TAB_Sgm}, Table~\ref{TAB_Sup}, Table~\ref{TAB_Sum} and Table~\ref{TAB_Pu}.

\item $\Pi_g$, $\Delta_g$, $\Delta_u$:
\\ $4 \times 4$ correlation matrices with operators as specified in Table~\ref{TAB_Pg}, Table~\ref{TAB_Dg} and Table~\ref{TAB_Du}.

\item $\Sigma_g^+$ (ground state):
\\ ordinary Wilson loops, i.e.\ $1 \times 1$ correlation matrices.

\item $\Sigma_g^+$ (first excitation):
\\ $3 \times 3$ correlation matrices with the same operators $S_j$ as for $\Sigma_u^+$ (cf.\ Table~\ref{TAB_Sup}).
\end{itemize}

We solve generalized eigenvalue problems
\begin{eqnarray}
C_{\Lambda_\eta^\epsilon}(r,t) \mathbf{v}^{(n)}(r,t,t_0) \ \ = \ \ \lambda^{(n)}(r,t,t_0) C_{\Lambda_\eta^\epsilon}(r,t_0) \mathbf{v}^{(n)}(r,t,t_0)
\end{eqnarray}
with $t_0 = a$ and $n = 0,1,\ldots$ (we sort the resulting eigenvalues according to $\lambda^{(0)}(r,t,t_0) > \lambda^{(1)}(r,t,t_0) > \ldots$; for details concerning the generalized eigenvalue problem cf.\ e.g.\ \cite{Blossier:2009kd} and references therein). The resulting ``effective potentials''
\begin{eqnarray}
V_{\textrm{eff};\Lambda_\eta^\epsilon}^{(n)}(r,t,t_0) \ \ = \ \ \ln \frac{\lambda^{(n)}(r,t,t_0)}{\lambda^{(n)}(r,t+a,t_0)}
\end{eqnarray}
are constant with respect to $t$ for sufficiently large $t$ within statistical errors. The plateau values of $V_{\textrm{eff};\Lambda_\eta^\epsilon}^{(0)}(r,t,t_0)$ correspond to the ground state potentials $V_{\Lambda_\eta^\epsilon}(r)$ and we extract them by fitting a constant to $V_{\textrm{eff};\Lambda_\eta^\epsilon}^{(0)}(r,t,t_0)$ for each $r$ in the range $t_\textrm{min} \leq t \leq t_\textrm{max}$. Similarly, we determine the first excitation $V'_{\Sigma_g^+}(r)$ from $V_{\textrm{eff};\Lambda_\eta^\epsilon}^{(1)}(r,t,t_0)$. In principle we could determine the first excitations of the hybrid static potentials in the same way, but since statistical errors for these excitations are quite large, we decided not to include the corresponding results in this work.

We choose $t_\textrm{min}$ sufficiently large to guarantee a strong suppression of excited states. On the other hand, statistical errors of effective potentials are smaller at smaller $t$. Consequently, the statistical error on $V_{\Lambda_\eta^\epsilon}(r)$ will be smaller, when using smaller $t_\textrm{min}$. We have implemented the following algorithm with the intention to automatically determine $t_\textrm{min}$ and $t_\textrm{max}$ for each of the static potentials and each $r$ in a fair way.
\begin{itemize}
\item $t'_\textrm{min}$ is the minimal $t$, where $V_{\textrm{eff};\Lambda_\eta^\epsilon}(r,t,t_0)$ and $V_{\textrm{eff};\Lambda_\eta^\epsilon}(r,t+a,t_0)$ differ by less than $2 \, \sigma$.

\item $t'_\textrm{max} = 9 \, a$, the maximum $t$, where correlation functions have been computed.

\item Fit constants $V_{\Lambda_\eta^\epsilon}(r)$ to $V_{\textrm{eff};\Lambda_\eta^\epsilon}(r,t,t_0)$ for all ranges $t_\textrm{min} \ldots t_\textrm{max}$ with $t'_\textrm{min} \leq t_\textrm{min}$, $t_\textrm{max} \leq t'_\textrm{max}$ and $t_\textrm{max} - t_\textrm{min} \geq 2 \, a$. Results with $\chi^2_\textrm{red} > 1.0$ are discarded, where $\chi^2_\textrm{red}$ denotes the uncorrelated reduced $\chi^2$ of the corresponding fit. If all fits yield $\chi^2_\textrm{red} > 1.0$, keep that one with the smallest $\chi^2_\textrm{red}$ and discard all others.

\item As final result for $V_{\Lambda_\eta^\epsilon}(r)$ take the fit result corresponding to the longest plateau, i.e.\ with maximum $t_\textrm{max} - t_\textrm{min}$. If there are several fit results with the same maximum $t_\textrm{max} - t_\textrm{min}$, take the fit result with the smallest $t_\textrm{min}$.
\end{itemize}
We have checked each of the resulting fitting ranges and corresponding plateau fits and have found that in almost all cases the algorithm decided for a reasonable range $t_\textrm{min} \ldots t_\textrm{max}$. Only in very few cases (less than 5\%) one gets the impression that there is a slight mismatch between the range $t_\textrm{min} \ldots t_\textrm{max}$ determined by the algorithm and the plateau region. In these cases we have changed $t_\textrm{min}$ manually by either $+a$ or $-a$.

\begin{figure}[p]
\begin{center}
\includegraphics[scale=0.68]{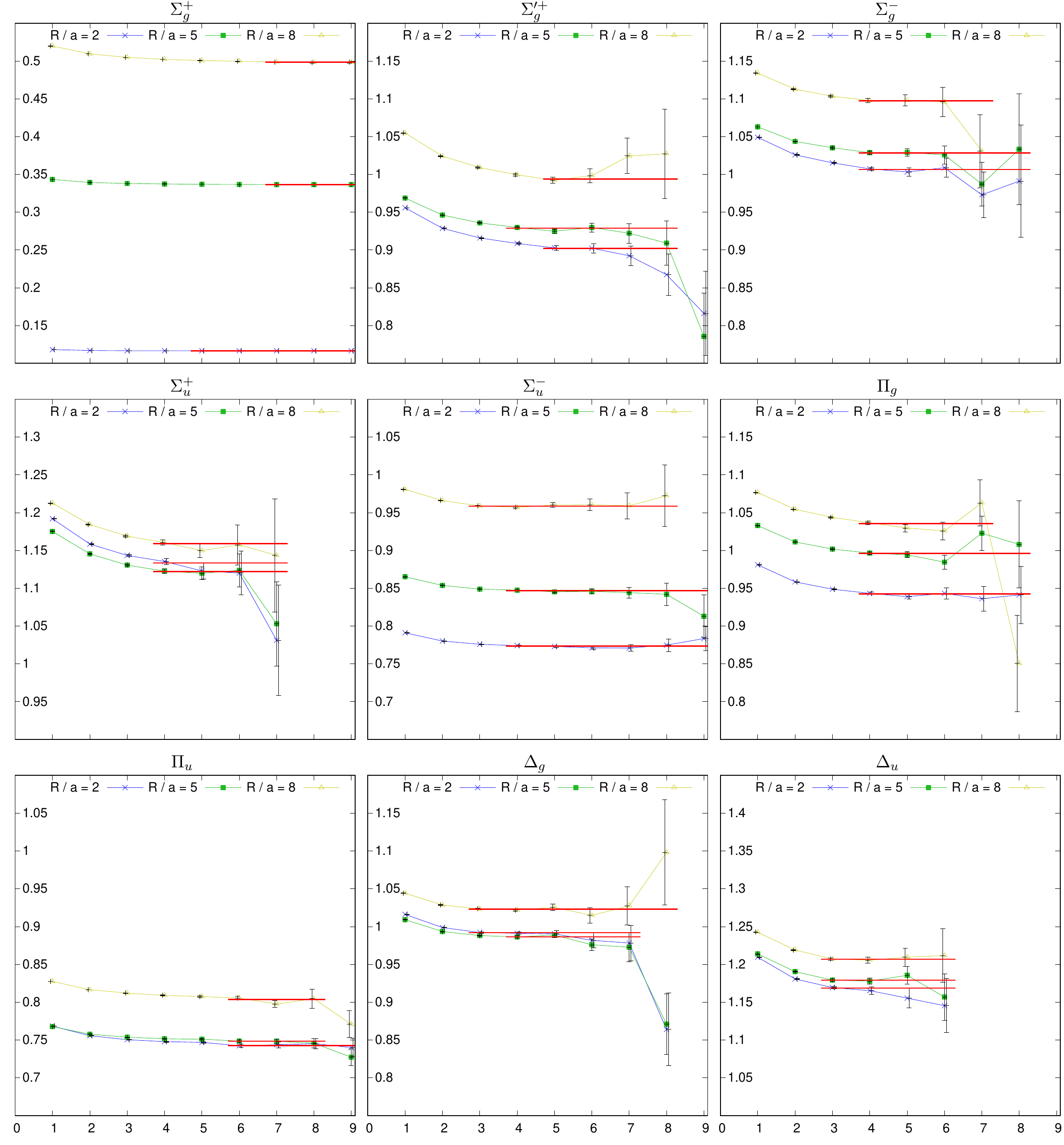}
\end{center}
\caption{\label{FIG001}Effective potentials $V_{\textrm{eff};\Lambda_\eta^\epsilon}^{(0)}(r,t,t_0=a) a$, $\Lambda_\eta^\epsilon = \Sigma_g^+,\Sigma_g^-,\Sigma_u^+,\Sigma_u^-,\Pi_g,\Pi_u,\Delta_g,\Delta_u$ and $V_{\textrm{eff};\Sigma_g^+}^{(1)}(r,t,t_0=a) a$ as functions of $t/a$ together with the corresponding plateau fits for separations $r/a = 2, 5, 8$. To allow a straightforward comparison of different sectors, the vertical scale is the same for all nine plots.}
\end{figure}

To illustrate the quality of our numerical data and the automatic determination of fitting ranges, we show effective potentials for all eight sectors $\Lambda_\eta^\epsilon = \Sigma_g^+,\Sigma_g^-,\Sigma_u^+,\Sigma_u^-,\Pi_g,\Pi_u,\Delta_g,\Delta_u$ for separations $r/a = 2, 5, 8$ in Figure~\ref{FIG001}. The fitting ranges $t_\textrm{min} \ldots t_\textrm{max}$ are indicated by red lines.

\begin{figure}[p]
\begin{center}
\includegraphics[scale=0.82]{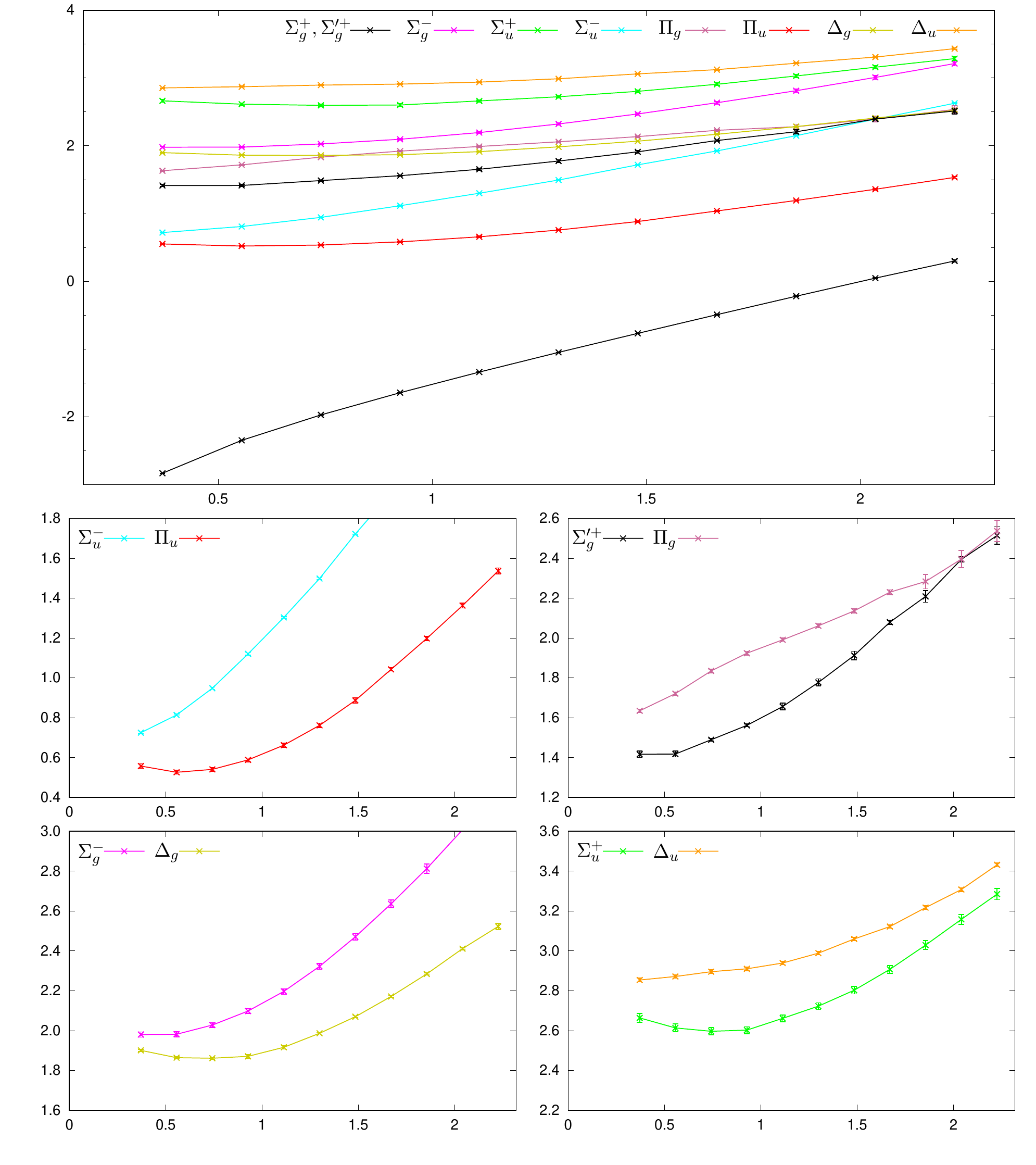}
\end{center}
\caption{\label{FIG002}The ordinary static potential $V_{\Sigma_g^+}(r) r_0$ and the corresponding first excitation $V'_{\Sigma_g^+}(r) r_0$ as well as the hybrid static potentials $V_{\Lambda_\eta^\epsilon}(r) r_0$, $\Lambda_\eta^\epsilon = \Sigma_g^-,\Sigma_u^+,\Sigma_u^-,\Pi_g,\Pi_u,\Delta_g,\Delta_u$ as functions of the separation $r / r_0$, where $r_0 = 0.5 \, \textrm{fm}$. To allow a straightforward comparison with results from the literature, e.g.\ with \cite{Juge:1997nc,Juge:2002br}, the vertical scale has been shifted by an additive constant such that $V_{\Sigma_g^+}(2 r_0) = 0$.}
\end{figure}

The resulting hybrid static potentials are shown in units of $r_0$ \footnote{We have plotted both the quark-antiquark separation $r$ as well as the static potentials $V_{\Lambda_\eta^\epsilon}(r)$ in units of $r_0 = 0.5 \, \textrm{fm}$ (cf.\ also section~\ref{SEC679}), to allow a straightforward comparison with the results from \cite{Juge:1997nc,Juge:2002br}, which are frequently used in recent publications.} together with the ordinary static potential and its first excitation in Figure~\ref{FIG002}. Separations $r < 2 a$ are not shown, because such data points are known to exhibit strong lattice discretization effects. The corresponding numbers $V_{\Lambda_\eta^\epsilon}(r)$, $\Lambda_\eta^\epsilon = \Sigma_g^+,\Sigma_g^-,\Sigma_u^+,\Sigma_u^-,\Pi_g,\Pi_u,\Delta_g,\Delta_u$ and $V'_{\Sigma_g^+}(r)$ are collected in Table~\ref{TAB400} for future reference. This data might be of interest for similar recent or future lattice studies as a benchmark (cf.\ e.g.\ \cite{Bicudo:2018jbb}) or as input for effective field theories like pNRQCD and mass determinations of heavy hybrid mesons (cf.\ e.g.\ \cite{Berwein:2015vca,Oncala:2017hop}).

In the following we compare our results from Figure~\ref{FIG001}, Figure~\ref{FIG002} and Table~\ref{TAB400} to a previous computation of hybrid static potentials \cite{Juge:1997ir,Morningstar:1998xh,Juge:1999ie,
Juge:1999aw,Morningstar:2001nu,Juge:2002br,
Juge:2003qd,Juge:2003ge}. Even though this computation was performed more than 15 years ago, the resulting potentials are frequently used in current projects (cf.\ e.g.\ \cite{Berwein:2015vca,Oncala:2017hop}) and seem to be the most accurate lattice field theory results for hybrid static potentials, which are currently available. Unfortunately, the above references correspond to rather short publications, where the resulting potentials are shown, but details are missing, e.g.\ tables containing numerical values, in particular error bars, or effective mass plots. More detailed information is, however, available on the webpage \cite{wwwCM} of one of the authors of the listed publications. Thus, the following comparative discussion is to a large extent based on the data collected at \cite{wwwCM}.

Both our computation as well as the computation from \cite{wwwCM} are done within pure SU(3) gauge theory, i.e.\ without dynamical quarks. There are, however, several technical differences.
\begin{itemize}
\item At \cite{wwwCM} different lattice spacings are considered, where the two finest spatial lattice spacings are $a \approx 0.12 \, \textrm{fm}$ (denoted as \textit{Run~A}) and $a \approx 0.19 \, \textrm{fm}$ (denoted as \textit{Run~B}). Our computation is done at a single lattice spacing, which is somewhat smaller, $a \approx 0.093 \, \textrm{fm}$.

\item The maximal separations provided at \cite{wwwCM} are $r = 12 \, a \approx 1.44 \, \textrm{fm}$ (\textit{Run~A}) and $r = 10 \, a \approx 1.90 \, \textrm{fm}$ (\textit{Run~B}). In our computation we considered separations up to $r = 12 \, a \approx 1.12 \, \textrm{fm}$.

\item We use the ordinary Wilson gauge action, i.e.\ lattices with the same lattice spacing in all four spacetime dimensions. At \cite{wwwCM} anisotropic lattices are employed, where the temporal lattice spacing is smaller by a factor of $3$ (\textit{Run~A}) and $5$ (\textit{Run~B}) than the spatial lattice spacing.

\item We use a $24^3 \times 48$ lattice, i.e.\ the same number of lattice sites in all three spatial directions. At \cite{wwwCM} the number of spatial lattice sites is larger in the direction of the quark antiquark separation, than in the two other directions, i.e.\ $(18^2 \times 24) \times 54$ (\textit{Run~A}) and $(16^2 \times 20) \times 80$ (\textit{Run~B}).

\item We use HYP2 smeared temporal links, when computing correlation functions (\ref{EQN600}), to reduce the self energy of the static quarks. At \cite{wwwCM} unsmeared temporal links are used.
\end{itemize}

Comparing our effective potentials (Figure~\ref{FIG001}) to those at \cite{wwwCM} in a meaningful way is difficult. Our effective potentials are obtained from correlation matrices by solving generalized eigenvalue problems, while at \cite{wwwCM} optimized correlation functions are used. Also the extraction of the potentials is done in a different way. We fit constants at larger temporal separations, where effective potentials are consistent with a plateau, while at \cite{wwwCM} sums of two exponentials are fitted to the optimized correlation functions including also data points at small temporal separations. Nevertheless, when comparing to \cite{wwwCM}/\textit{Run A} it seems that our effective potentials start to be consistent with plateaus at smaller temporal separations (in physical units) for $\Lambda_\eta^\epsilon = \Sigma_u^-, \Delta_g, \Delta_u$, at similar temporal separations for $\Lambda_\eta^\epsilon = \Sigma_u^+, \Sigma_g^-, \Pi_g$ and at larger temporal separations for $\Lambda_\eta^\epsilon = \Pi_u$. This is most likely a consequence of different operator sets used in this work and at \cite{wwwCM}. While we have documented our operator optimization in section~\ref{SEC445} in detail, equivalent information for the computation from \cite{wwwCM} does not seem to be available. Another observation is that our plateau-like regions tend to be somewhat longer (in physical units), which we attribute to the smaller self energy of the static quarks due to the use of HYP2 smeared temporal links.

The temporal separations, where effective potentials start to be consistent with plateaus, are also reflected in the statistical errors of the hybrid static potentials (cf.\ Table~\ref{TAB400} of this work and \cite{wwwCM}). In comparison to \cite{wwwCM}/\textit{Run A} our statistical errors are smaller by a factor $\approx 1.5 \ldots 2.0$ for $\Lambda_\eta^\epsilon = \Sigma_u^-, \Delta_g, \Delta_u$, similar for $\Lambda_\eta^\epsilon = \Sigma_u^+, \Sigma_g^-, \Pi_g$ and larger by a factor $\approx 2.0$ for $\Lambda_\eta^\epsilon = \Pi_u$. \cite{wwwCM}/\textit{Run B} has slightly smaller statistical errors than \cite{wwwCM}/\textit{Run A}, but the overall picture is the same.

There is no obvious discrepancy for the majority of potentials concerning their shape (cf.\ Figure~\ref{FIG002} of this work and e.g.\ FIG.~2 in \cite{Juge:2002br}). Clearly visible differences can be observed for $V_{\Pi_g}(r)$ and $V_{\Delta_u}(r)$, in particular  at small separations $r \ltapprox 0.25 \, \textrm{fm}$. Our results for these potentials are somewhat lower than those from \cite{Juge:2002br} and exhibit the expected approximate degeneracy with $V'_{\Sigma^+_g}(r)$ and $V_{\Sigma_u^+}(r)$, respectively (for a detailed discussion of these degeneracies and their relation to gluelump masses cf.\ e.g.\ \cite{Brambilla:1999xf}). Interestingly, we have found that the resulting potentials $V_{\Pi_g}(r)$ and $V_{\Delta_u}(r)$ are quite sensitive to the creation operators used in the correlation matrices. In both cases the operator $S_{IV,2}$ significantly increases the ground state overlap and, thus, is essential to observe the previously mentioned and expected degeneracies at short $r$. We interpret this as indication that our selected sets of operators are better able to isolate the groundstate potentials for short $r$ in the $\Pi_g$ and $\Delta_u$ sectors than the operators used in \cite{Juge:2002br}. It should, however, be noted that hybrid static potentials at sufficiently small separations can decay to the ordinary static potential and a glueball. Thus, extracting the potential from the exponential decay of a correlation function, as done in our work as well as in \cite{Juge:2002br}, might give contaminated results for small $r$. The lightest glueball has quantum numbers $J^{P C} = 0^{++}$ and mass $m_{0^{++}} \approx 4.21 / r_0$ \cite{Morningstar:1999rf}. Using this mass one can read off from Figure~\ref{FIG002} that $V_{\Pi_g}(r)$ can decay for $r \ltapprox 0.25 \, \textrm{fm}$ and $V_{\Delta_u}(r)$ for $r \ltapprox 0.5 \, \textrm{fm}$. The lowest hybrid static potentials $V_{\Pi_u}(r)$ and $V_{\Sigma_u^-}(r)$, which are used in section~\ref{SEC498} to estimate masses of heavy hybrid mesons, can only decay for $r \ltapprox 0.12 \, \textrm{fm}$.

There are further lattice field theory computations of hybrid static potentials, which are interesting to discuss or to compare with:
\begin{itemize}
\item In \cite{Bali:2003jq} the $\Sigma_u^-$ and $\Pi_u$ hybrid static potentials were computed in pure SU(3) gauge theory using several lattice spacings $a \geq 0.16 \times r_0 \approx 0.08 \, \textrm{fm}$ as well as off-axis separations. The focus of the paper is on the phenomenology of static sources and gluonic excitations at short separation. The results presented for the $\Sigma_u^-$ and $\Pi_u$ hybrid static potentials seem to agree with our findings.

\item Very recently color field densities of static potential flux tubes in the sectors $\Sigma_g^+$, $\Sigma_u^+$ and $\Pi_u$ have been computed in pure SU(3) gauge theory \cite{Bicudo:2018jbb}. As a byproduct the potentials in these three sectors have been obtained, which seem to agree with our results.
\end{itemize}

\begin{table}[p]
\begin{center}

\def\arraystretch{1.2}

\begin{tabular}{cccccc}
\midrule
$r/a$ & $V_{\Sigma^+_g}a$ & $V'_{\Sigma^+_g}a$ & $V_{\Sigma^-_g}a$ & $V_{\Sigma^+_u}a$ & $V_{\Sigma^-_u}a$ \\ \midrule
2 & 0.116648(13) 			& 0.9020(30) & 1.0066(25) & 1.1334(42) & 0.77365(75) \\
3 & 0.206462(31) 			& 0.9023(28) & 1.0068(23) & 1.1241(37) & 0.79015(81) \\
4 & 0.275767(61) 			& 0.9155(14) & 1.0155(23) & 1.1209(35) & 0.81509(90) \\
5 & 0.33655(12)\phantom{0} & 0.9289(14) & 1.0285(23) & 1.1220(33) & 0.8469(11)\phantom{0} \\
6 & 0.39290(19)\phantom{0} & 0.9465(31) & 1.0468(25) & 1.1329(32) & 0.88094(63) \\
7 & 0.44651(29)\phantom{0} & 0.9689(34) & 1.0700(27) & 1.1443(33) & 0.91711(72) \\
8 & 0.49847(45)\phantom{0} & 0.9940(40) & 1.0975(31) & 1.1592(35) & 0.95862(81) \\
9 & 0.54952(68)\phantom{0} & 1.0249(19) & 1.1282(36) & 1.1786(37) & 0.99662(94) \\
10 & 0.6000(11)\phantom{00} & 1.0489(55) & 1.1610(43) & 1.2011(42) & 1.0382(11)\phantom{0} \\
11 & 0.6492(16)\phantom{00} & 1.0834(26) & 1.1971(53) & 1.2251(46) & 1.0831(13)\phantom{0} \\
12 & 0.6962(24)\phantom{00} & 1.1056(81) & 1.2350(64) & 1.2486(52) & 1.1266(15)\phantom{0} \\
\midrule
\end{tabular}

\begin{tabular}{ccccc}
\midrule
$r/a$ & $V_{\Pi_g}a$ & $V_{\Pi_u}a$ & $V_{\Delta_g}a$ & $V_{\Delta_u}a$ \\ \midrule
2 & 0.9425(17) & 0.7427(22) & 0.99183(88) 			& 1.1686(18) \\
3 & 0.9585(17) & 0.7369(19) & 0.98505(83) 			& 1.1719(17) \\
4 & 0.9796(18) & 0.7395(18) & 0.98451(78) 			& 1.1764(17) \\
5 & 0.9960(19) & 0.7483(18) & 0.9863(17)\phantom{0}& 1.1791(17) \\
6 & 1.0086(19) & 0.7621(20) & 0.9947(17)\phantom{0}& 1.1844(16) \\
7 & 1.0216(20) & 0.7805(22) & 1.00777(78) 			& 1.1936(16) \\
8 & 1.0355(21) & 0.8037(25) & 1.02320(81) 			& 1.2068(16) \\
9 & 1.0528(24) & 0.8326(16)	 & 1.04199(86)			& 1.2184(16) \\
10 & 1.0628(68) & 0.8613(19) & 1.06294(92)			& 1.2361(17) \\
11 & 1.0837(81) & 0.8920(23) & 1.0865(11)\phantom{0}& 1.2528(18) \\
12 & 1.1098(99) & 0.9243(28) & 1.1072(28)\phantom{0}& 1.2758(19) \\
\midrule
\end{tabular}

\end{center}
\caption{\label{TAB400}Summary of lattice field theory results for static potentials $V_{\Lambda_\eta^\epsilon}(r)$ with $\Lambda_\eta^\epsilon = \Sigma_g^+,\Sigma_g^-,\Sigma_u^+,\Sigma_u^-,\Pi_g,\Pi_u,\Delta_g,\Delta_u$ and $V'_{\Sigma_g^+}$.}
\end{table}


\newpage

\section{\label{SEC498}Masses of heavy hybrid mesons in the Born-Oppenheimer approximation}

In this section we parameterize the ordinary static potential $V_{\Sigma_g^+}(r)$ and the two lowest hybrid static potentials $V_{\Pi_u}(r)$ and $V_{\Sigma_u^-}(r)$ computed in section~\ref{SEC499}, to estimate masses of heavy hybrid mesons with quarks $\bar{Q} Q = \bar{c} c$ and $\bar{Q} Q = \bar{b} b$ for various $J^{P C}$ quantum numbers. This is done in the Born-Oppenheimer approximation \cite{bo}, which is a two-step procedure commonly used e.g.\ in molecular physics. In the first step, which is the computation of hybrid static potentials using lattice field theory (cf.\ sections \ref{SEC499} and \ref{SEC679}), the gluons are the only dynamical degrees of freedom, whereas the positions of the heavy quarks $\bar{Q}$ and $Q$ are frozen. In the second step, which is discussed in section~\ref{SEC755}, this constraint is relaxed by solving the Schr\"odinger equation for the relative coordinate of the $\bar{Q} Q$ pair using the hybrid static potentials computed in the first step.


\subsection{\label{SEC679}Parameterization of lattice field theory results for the ordinary $\Sigma_g^+$ and the hybrid $\Pi_u$ and $\Sigma_u^-$ static potentials}

For the ordinary static potential (the ground state in the $\Sigma_g^+$ sector) a common choice, which is able to parameterize lattice data for separations $r \gtapprox 0.15 \, \textrm{fm}$, is
\begin{eqnarray}
\label{EQN952} V_{\Sigma_g^+}(r) \ \ = \ \ V_0 - \frac{\alpha}{r} + \sigma r .
\end{eqnarray}
$\sigma$ is the string tension, $\alpha$ is a positive constant and $V_0$ is a physically irrelevant shift, which contains the self energy of the static quarks and, thus, depends on the lattice spacing. For a detailed recent discussion of this parameterization cf.\ e.g.\ \cite{Karbstein:2018mzo}. 

In this section we focus on the two lowest hybrid static potentials with quantum numbers $\Pi_u$ and $\Sigma_u^-$. Our parameterizations are based on the pNRQCD prediction
\begin{eqnarray}
\label{EQN340} V_\textrm{hybrid}(r) \ \ = \ \ V_o^\textrm{RS}(r,\nu_f) + \Lambda_H(\nu_f) + \mathcal{O}(r^2) ,
\end{eqnarray}
which is valid for small separations $r \ll 1 / \Lambda_\textrm{QCD} \approx 0.5 \, \textrm{fm}$ \cite{Brambilla:1999xf,Berwein:2015vca}. $V_o^\textrm{RS}(r,\nu_f) = \alpha_{V_0}(\nu_f) / 6 r + \delta V_o^\textrm{RS}(\nu_f)$ is the Renormalon Subtracted (RS) octet potential, $\nu_f$ the subtraction scale and $\Lambda_H(\nu_f)$ a constant. $\Lambda_H(\nu_f)$ is the same for those hybrid static potentials, which are degenerate for $r \rightarrow 0$, i.e.\ for $V_{\Pi_u}(r)$ and $V_{\Sigma_u^-}(r)$. Eq.\ (\ref{EQN340}) suggests to use
\begin{eqnarray}
\label{EQN401} V_{\Lambda_\eta^\epsilon}(r) \ \ = \ \ \frac{A_1}{r} + A_2 + A_3 r^2
\end{eqnarray}
as fit function for small separations $r$, where both $A_1$ und $A_2$ are the same for $\Lambda_\eta^\epsilon = \Pi_u$ and $\Lambda_\eta^\epsilon = \Sigma_u^-$, while $A_3$ is different. 

We are interested in parameterizations of our lattice field theory results over the whole available range of separations $r$, i.e.\ up to $r = 12 a \approx 1.1 \, \textrm{fm}$. While eq.\ (\ref{EQN401}) is suited to parameterize $V_{\Pi_u}(r)$ up to $r = 12 a$, which is beyond the region of validity of the pNRQCD prediction, this is not the case for $V_{\Sigma_u^-}(r)$. Therefore, we use for $V_{\Sigma_u^-}(r)$ a fit function with additional degrees of freedom, which reduces to eq.\ (\ref{EQN401}) in the limit of small $r$. In \cite{Braaten:2014qka} it was suggested to use
\begin{eqnarray}
\label{EQN953} V_{\Sigma_u^-}(r) - V_{\Pi_u}(r) \ \ = \ \ \frac{B_1 r^2}{1 + B_3 r^2}
\end{eqnarray}
for the difference of the two lowest hybrid static potentials. While this is a reasonable crude description of this difference, it is not sufficient to parameterize our precise lattice field theory results from section~\ref{SEC499} in a consistent way, i.e.\ with reduced $\chi^2 \ltapprox 1$. Therefore, we extend eq.\ (\ref{EQN953}) by introducing another parameter $B_2$,
\begin{eqnarray}
V_{\Sigma_u^-}(r) - V_{\Pi_u}(r) \ \ = \ \ \frac{B_1 r^2}{1 + B_2 r + B_3 r^2} .
\end{eqnarray}
Altogether we parameterize $V_{\Pi_u}(r)$ and $V_{\Sigma_u^-}(r)$ by
\begin{eqnarray}
\label{EQN954} & & \hspace{-0.7cm} V_{\Pi_u}(r) \ \ = \ \ \frac{A_1}{r} + A_2 + A_3 r^2 \\
\label{EQN955} & & \hspace{-0.7cm} V_{\Sigma_u^-}(r) \ \ = \ \ \frac{A_1}{r} + A_2 + A_3 r^2 + \frac{B_1 r^2}{1 + B_2 r + B_3 r^2} ,
\end{eqnarray}
where $A_1$, $A_2$, $A_3$, $B_1$, $B_2$ and $B_3$ are fit parameters. Note that (\ref{EQN955}) still reduces to the NRQCD prediction (\ref{EQN401}) in the limit of small $r$, i.e.\ the $\Pi_u$ and $\Sigma_u^-$ hybrid static potentials are dominated by the same repulsive ``octet-like'' $1/r$ term and become degenerate ($A_3$ in eq.\ (\ref{EQN401}) is then equivalent to $A_3 + B_1$ in eq.\ (\ref{EQN955})).

To determine the unknown parameters in eq.\ (\ref{EQN952}), eq.\ (\ref{EQN954}) and eq.\ (\ref{EQN955}), we perform uncorrelated $\chi^2$ minimizing fits to our lattice data points in the region $r_\textrm{min} \leq r \leq r_\textrm{max}$.

\begin{figure}[htb]
\begin{center}
\includegraphics[width=7.8cm]{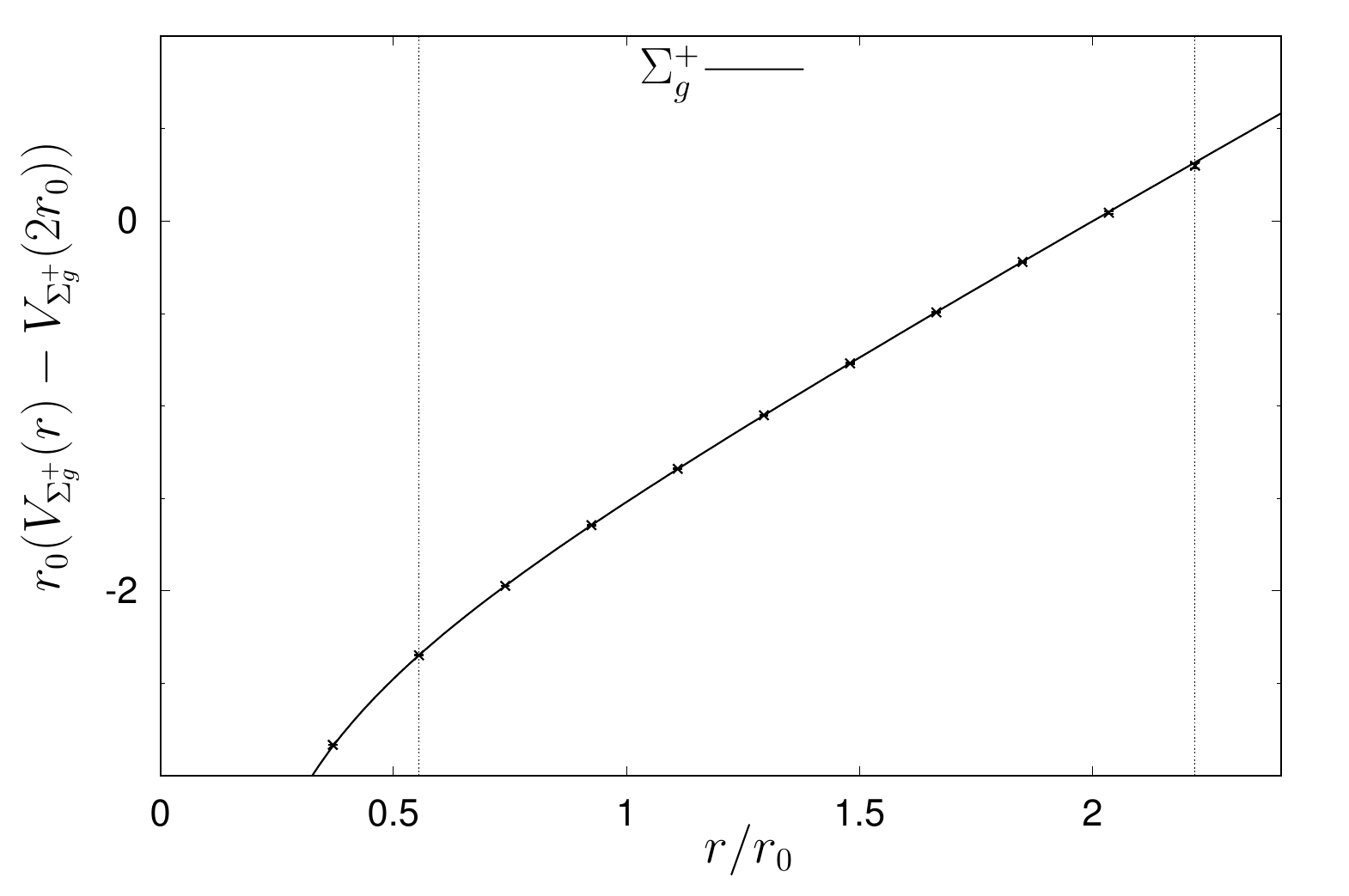}
\includegraphics[width=7.8cm]{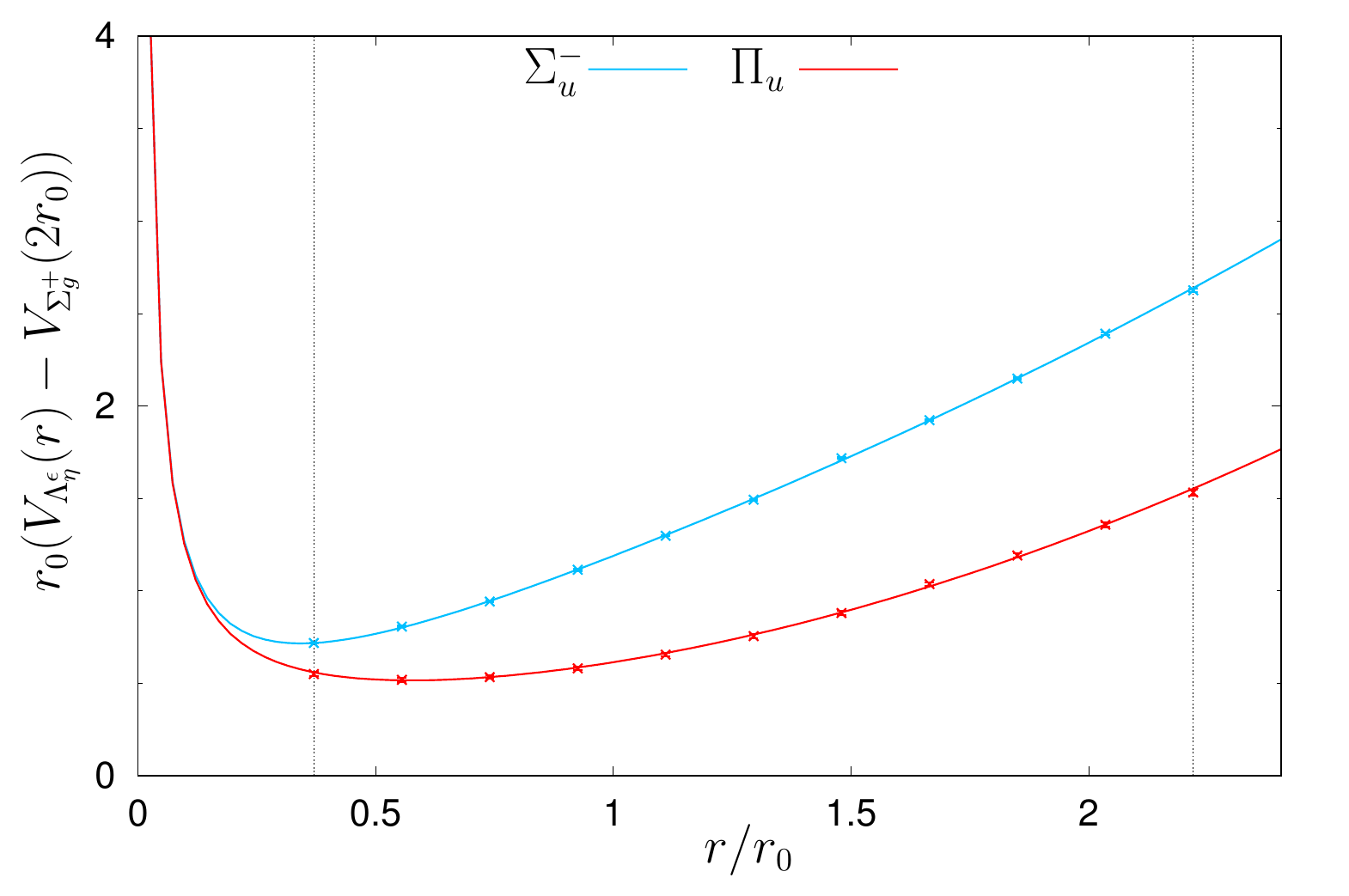}
\end{center}
\caption{\label{FIG105}Parameterizations of lattice field theory results. \textbf{(left)}~$\Sigma_g^+$ static potential (eq.\ (\ref{EQN952}) with parameters (\ref{EQN956})). \textbf{(right)}~$\Pi_u$ and $\Sigma_u^-$ hybrid static potentials (eq.\ (\ref{EQN954}) and (\ref{EQN955}) with parameters (\ref{EQN957})). Vertical dotted lines indicate the data points, which have been considered in the $\chi^2$ minimizing fits.}
\end{figure}

For the $\Sigma_g^+$ static potential we use $r_\textrm{min} = 3 a$ (for $r < 3 a$ lattice field theory results obtained with HYP smearing typically exhibit non-negligible discretization errors) and $r_\textrm{max} = 12 a$. From a 3-parameter fit we obtain
\begin{eqnarray}
\label{EQN956} \begin{array}{rrr}
V_0 a = 0.1515(13) & \alpha = 0.2626(23) & \sigma a^2 = 0.04749(17)
\end{array} ,
\end{eqnarray}
where $\chi_\textrm{red}^2 = 0.80$ indicates a consistent fit. The parameterization (\ref{EQN952}) with the parameters (\ref{EQN956}) is shown in Figure~\ref{FIG105} (left) together with our corresponding lattice field theory results.

From these results we can determine $r_0$, which is the typical length scale in lattice gauge theory and quite often used to set the scale. $r_0$ is defined via
\begin{eqnarray}
V'(r_0) r_0^2 \ \ = \ \ 1.65
\end{eqnarray}
and we obtain
\begin{eqnarray}
\frac{r_0}{a} \ \ = \ \ \bigg(\frac{1.65 - \alpha}{\sigma a^2}\bigg)^{1/2} \ \ = \ \ 5.405(6) .
\end{eqnarray}
When identifying $r_0$ with $0.5 \, \textrm{fm}$, which is a common choice in lattice gauge theory \footnote{For a discussion of the uncertainty of $r_0$ in $\textrm{fm}$ cf.\ e.g.\ \cite{Sommer:2014mea}, in particular Table~1.}, we find $a = 0.0925(1) \, \textrm{fm}$ (cf.\ also section~\ref{SEC567}). This value is consistent with an independent simulation and scale setting analysis quoted in \cite{Koma:2006si}.

For the $\Pi_u(r)$ and $\Sigma_u^-$ hybrid static potentials we use $r_\textrm{min} = 2 \, a$ and $r_\textrm{max} = 12 \, a$. From a single 6-parameter fit we obtain
\begin{eqnarray}
\label{EQN957} \begin{array}{rrr}
A_1 = 0.0958(46) & A_2 a = 0.6900(30) & A_3 a^3 = 0.001599(29) \\
B_1 a^3 = 0.0119(10) & B_2 a = 0.249(42)\phantom{0} & B_3 a^2 = 0.0316(28)\phantom{00}
\end{array} ,
\end{eqnarray}
where $\chi_\textrm{red}^2 = 1.48$ indicates a reasonable fit. The parameterizations (\ref{EQN954}) and (\ref{EQN955}) with the parameters (\ref{EQN957}) are shown in Figure~\ref{FIG105} (right) together with our corresponding lattice field theory results.


\subsection{\label{SEC755}Prediction of masses of heavy hybrid mesons}

The Born-Oppenheimer approximation for heavy hybrid mesons was pioneered in \cite{Hasenfratz:1980jv,Perantonis:1990dy,Juge:1999ie} and is explained in detail in \cite{Braaten:2014qka}. One has to solve the radial Schr\"odinger equation
\begin{eqnarray}
\label{EQN742} \bigg(-\frac{1}{2 \mu} \frac{d^2}{dr^2} + \frac{L (L+1) - 2 \Lambda^2 + J_{\Lambda_{\eta}^{\epsilon}} (J_{\Lambda_{\eta}^{\epsilon}}+1)}{2 \mu r^2} + V_{\Lambda_{\eta}^{\epsilon}}(r)\bigg) u_{\Lambda_{\eta}^{\epsilon};L,n}(r) \ \ = \ \ E_{\Lambda_{\eta}^{\epsilon};L,n} u_{\Lambda_{\eta}^{\epsilon};L,n}(r) ,
\end{eqnarray}
where $r$ is the separation of the heavy $\bar{Q} Q$ pair, $V_{\Lambda_{\eta}^{\epsilon}}(r)$ is one of the static potential parameterizations from section~\ref{SEC679}, eq.\ (\ref{EQN952}), (\ref{EQN954}) or (\ref{EQN955}), and $\mu = m_{\bar{Q}} m_Q /  (m_{\bar{Q}} + m_Q)$ is the reduced mass of the $Q \bar{Q}$ pair. We use $m_Q = m_c = 1628 \, \textrm{MeV}$ and $m_Q = m_b = 4977 \, \textrm{MeV}$ from quark models \cite{Godfrey:1985xj}. The wave function of the relative coordinate of the $\bar{Q} Q$ pair is $\psi_{\Lambda_{\eta}^{\epsilon};L,n,m_L}(r,\vartheta,\varphi) = (u_{\Lambda_{\eta}^{\epsilon};L,n}(r) / r) Y_{L,m_L}(\vartheta,\varphi)$. $L \in \{ \Lambda,\Lambda+1,\ldots \}$ is the quantum number corresponding to the operator $\mathbf{L}$, the sum of all angular momenta excluding the heavy quark spins $\mathbf{S}$, i.e.\ $\mathbf{J} = \mathbf{L} + \mathbf{S}$, where $\mathbf{J}$ is the total angular momentum of the meson. In the limit $r \rightarrow 0$ the gluon field configuration of a hybrid static potential is identical to that of a gluelump, where $J_{\Lambda_{\eta}^{\epsilon}}$ is the gluon spin of this gluelump. $J_{\Lambda_{\eta}^{\epsilon}} = 0$ for $\Lambda_{\eta}^{\epsilon} = \Sigma_g^+$, $J_{\Lambda_{\eta}^{\epsilon}} = 1$ for $\Lambda_{\eta}^{\epsilon} \in \{ \Sigma_g^+{}' , \Sigma_u^- , \Pi_g , \Pi_u \}$ and $J_{\Lambda_{\eta}^{\epsilon}} = 2$ for $\Lambda_{\eta}^{\epsilon} \in \{ \Sigma_g^- , \Sigma_u^+ , \Delta_g , \Delta_u \}$ \cite{Brambilla:1999xf}.

The derivation of the Schr\"odinger equation (\ref{EQN742}) is based on the following approximations (cf.\ also \cite{Braaten:2014qka}):
\begin{itemize}
\item In the \textit{adiabatic approximation} the gluon field is assumed to be in a stationary state in the presence of the heavy $\bar{Q} Q$ pair, i.e.\ the gluon field configuration is one of the hybrid static potentials computed in section~\ref{SEC499} labeled by quantum numbers $\Lambda_{\eta}^{\epsilon}$. Errors are proportional to $\Lambda_\textrm{QCD} / m_Q$, i.e.\ the adiabatic approximation is suited for heavy quarks . It is consistent with using static potentials, where also $1/m_Q$ corrections are neglected.

\item The Schr\"odinger equation to determine masses of a heavy hybrid mesons with quantum numbers $J^P$ is a multi-channel equation including all hybrid static potentials $V_{\Lambda_{\eta}^{\epsilon}}(r)$ consistent with $J^P$ (cf.\ \cite{Berwein:2015vca} for a detailed derivation of a coupled channel Schr\"odinger equation). In the \textit{single channel approximation} only a single component of this multi-channel Schr\"odinger equation is considered and couplings to other channels are ignored. The single channel approximation is good, if the resulting wave function is small for separations $r$, where the used hybrid static potential has avoided crossings with the other hybrid static potentials.

\item Finally the gluon spin is approximated by the gluon spin of a gluelump, which is a good approximation for small separations $r$, where the system resembles a gluelump. Consequently, the approximation is good for resulting wave functions $\psi(r,\vartheta,\varphi)$, which are localized near $r = 0$.
\end{itemize}

The Schr\"odinger equation (\ref{EQN742}) can be solved numerically with standard techniques. We employ a 4th order Runge-Kutta shooting method combined with Newton's method for root finding. Note that the resulting energies $E_{\Lambda_{\eta}^{\epsilon};L,n}$ contain the self-energies of the static quarks, which depend on the lattice spacing. To predict heavy hybrid meson masses, one has to eliminate these self-energies, which we do by subtracting $E_{\Lambda_{\eta}^{\epsilon}=\Sigma_g^+;n=1,L=0}$, the lowest energy from the ordinary static potential computed within the same setup. This energy $E_{\Lambda_{\eta}^{\epsilon}=\Sigma_g^+;n=1,L=0}$ corresponds for $\bar{Q} Q = \bar{c} c$ to the $\eta_c(1S)$ and $J/\Psi(1S)$ meson, which are degenerate in the static limit, and similarly for $\bar{Q} Q = \bar{b} b$ to the $\eta_b(1S)$ and $\Upsilon(1S)$ meson. Heavy hybrid meson masses are then given by
\begin{eqnarray}
\label{EQN864} m_{\Lambda_{\eta}^{\epsilon};L,n} \ \ = \ \ E_{\Lambda_{\eta}^{\epsilon};L,n} - E_{\Lambda_{\eta}^{\epsilon}=\Sigma_g^+;n=1,L=0} + \overline{m} ,
\end{eqnarray}
where $\overline{m}$ is the spin averaged mass from experiments, either $\overline{m} = (m_{\eta_b(1S),\textrm{exp}} + 3 m_{\Upsilon(1S),\textrm{exp}}) / 4 = 9445(1) \, \textrm{MeV}$ or $\overline{m} = (m_{\eta_c(1S),\textrm{exp}} + 3 m_{J/\Psi(1S),\textrm{exp}}) / 4 = 3069(1) \, \textrm{MeV}$ \cite{Tanabashi:2018oca}.



\begin{figure}[htb]
\begin{center}
$\bar{c} c$ hybrid meson masses \\
\includegraphics[width=7.8cm]{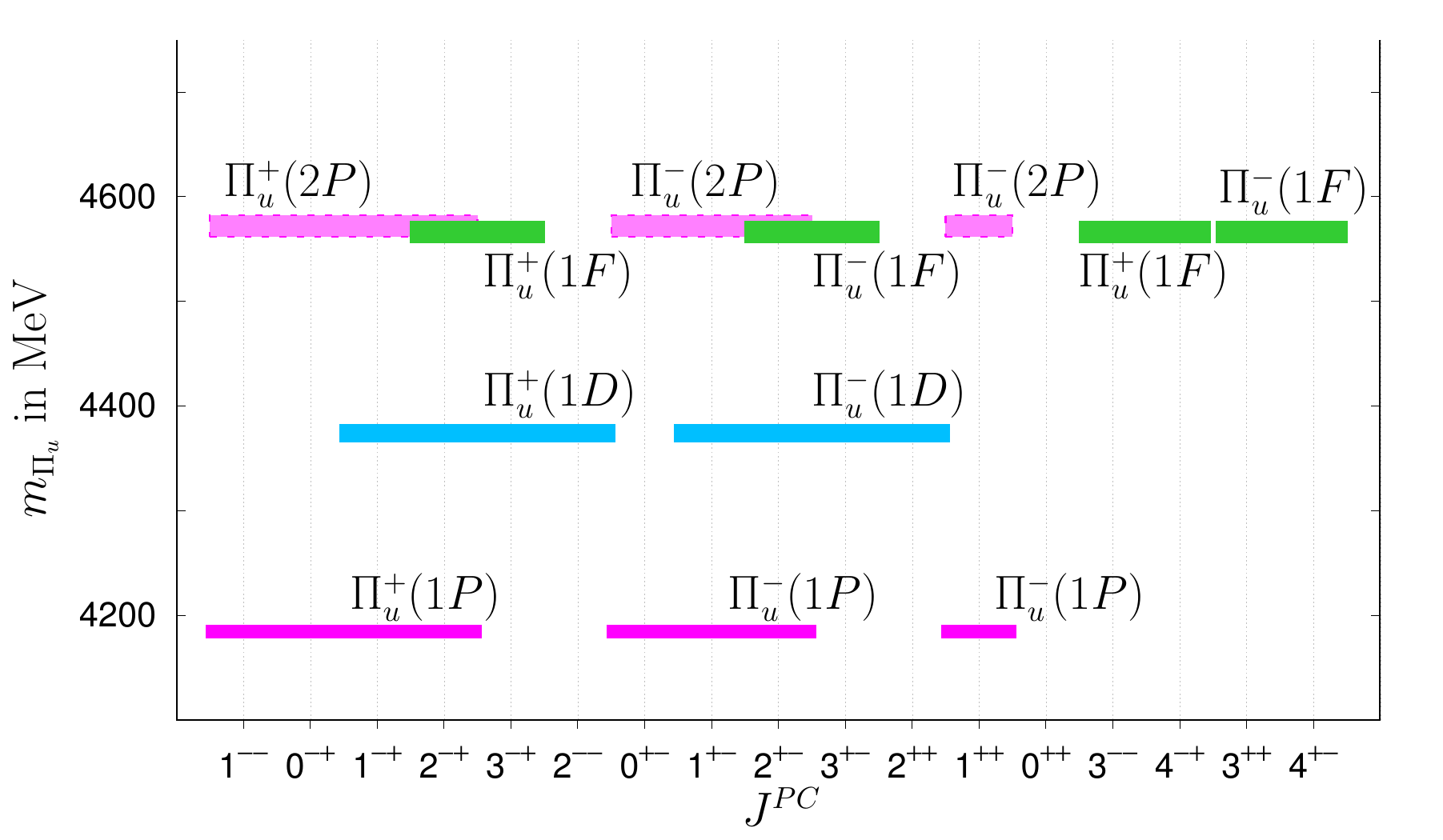}
\includegraphics[width=7.8cm]{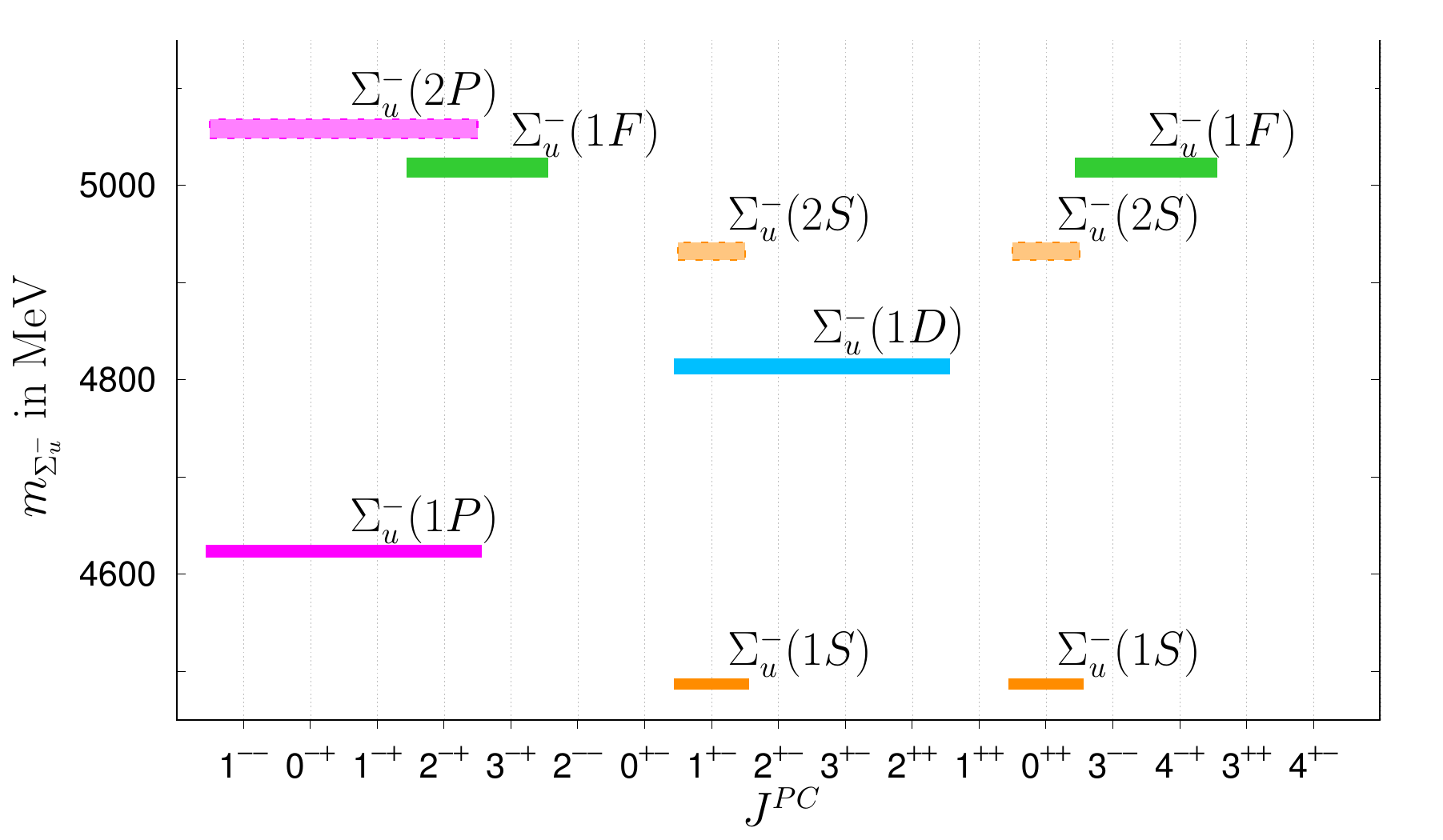}

\vspace{0.2cm}
$\bar{b} b$ hybrid meson masses \\
\includegraphics[width=7.8cm]{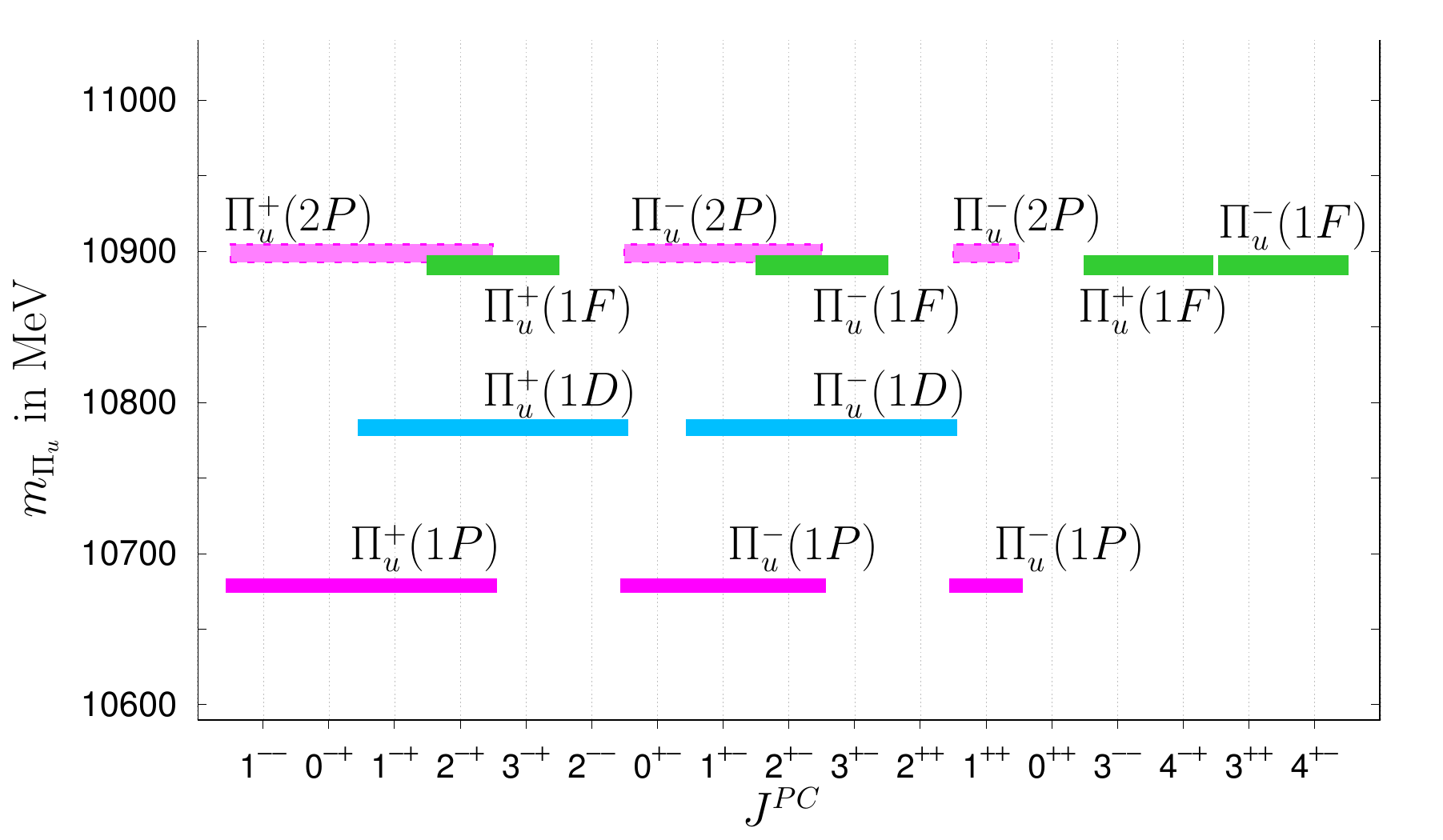}
\includegraphics[width=7.8cm]{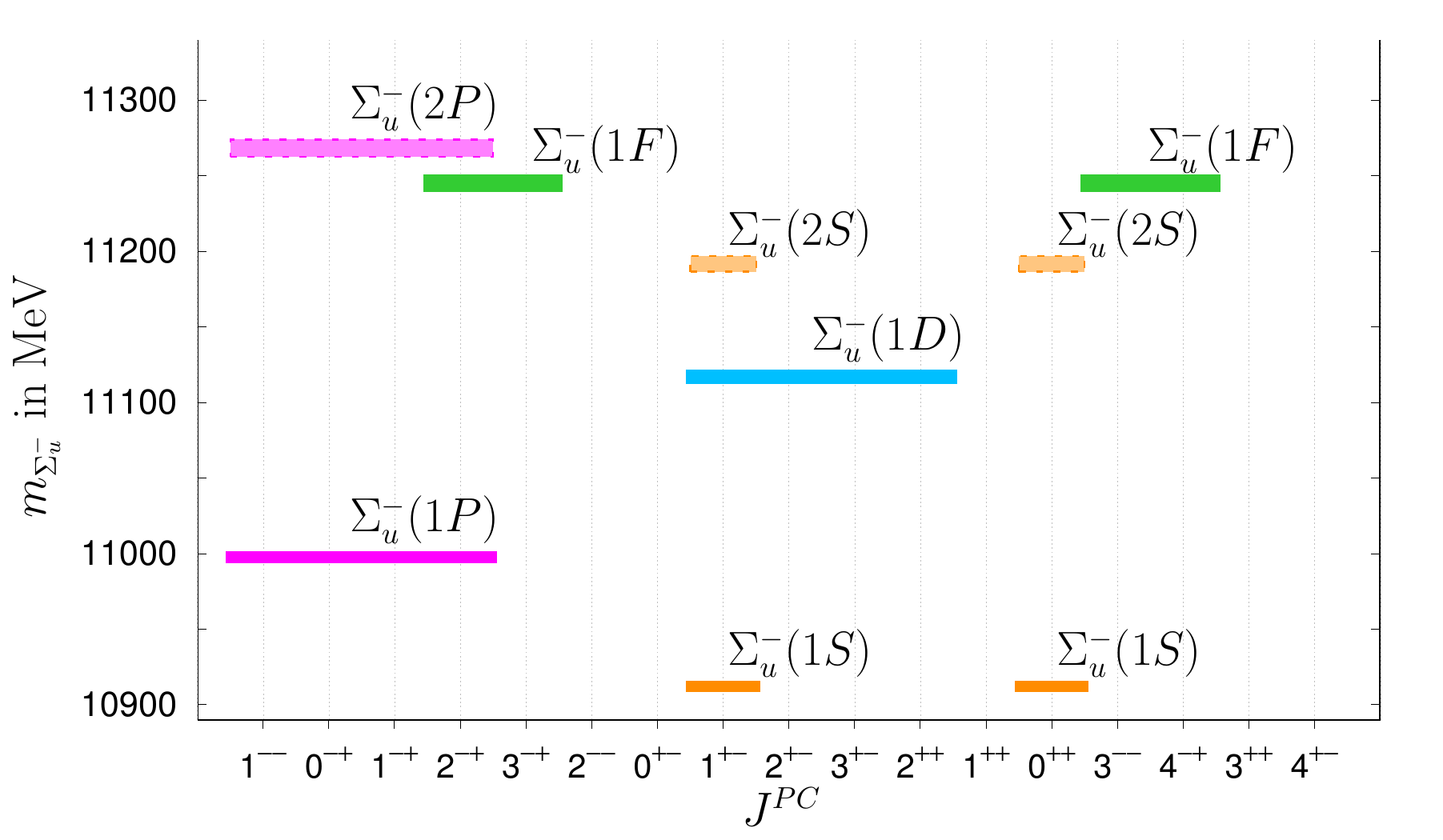}
\end{center}
\caption{\label{FIG831}Predictions for heavy hybrid meson masses. \textbf{(left)}~$\Pi_u$ hybrid static potential. \textbf{(right)}~$\Sigma_u^-$ hybrid static potential.}
\end{figure}

\begin{figure}[htb]
\begin{center}
radial probability density for $\bar{c} c$ hybrid mesons \\
\includegraphics[width=7.8cm]{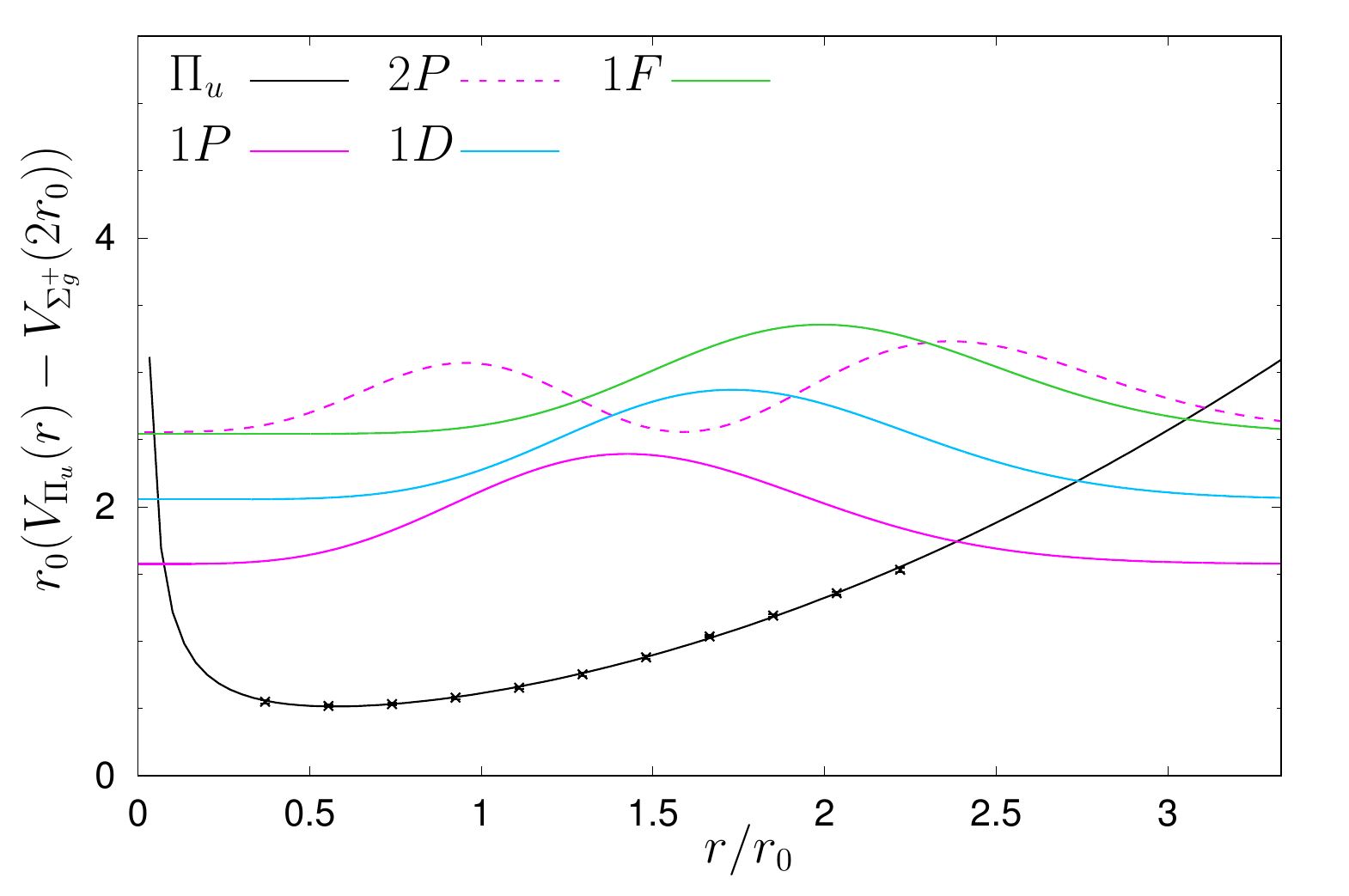}
\includegraphics[width=7.8cm]{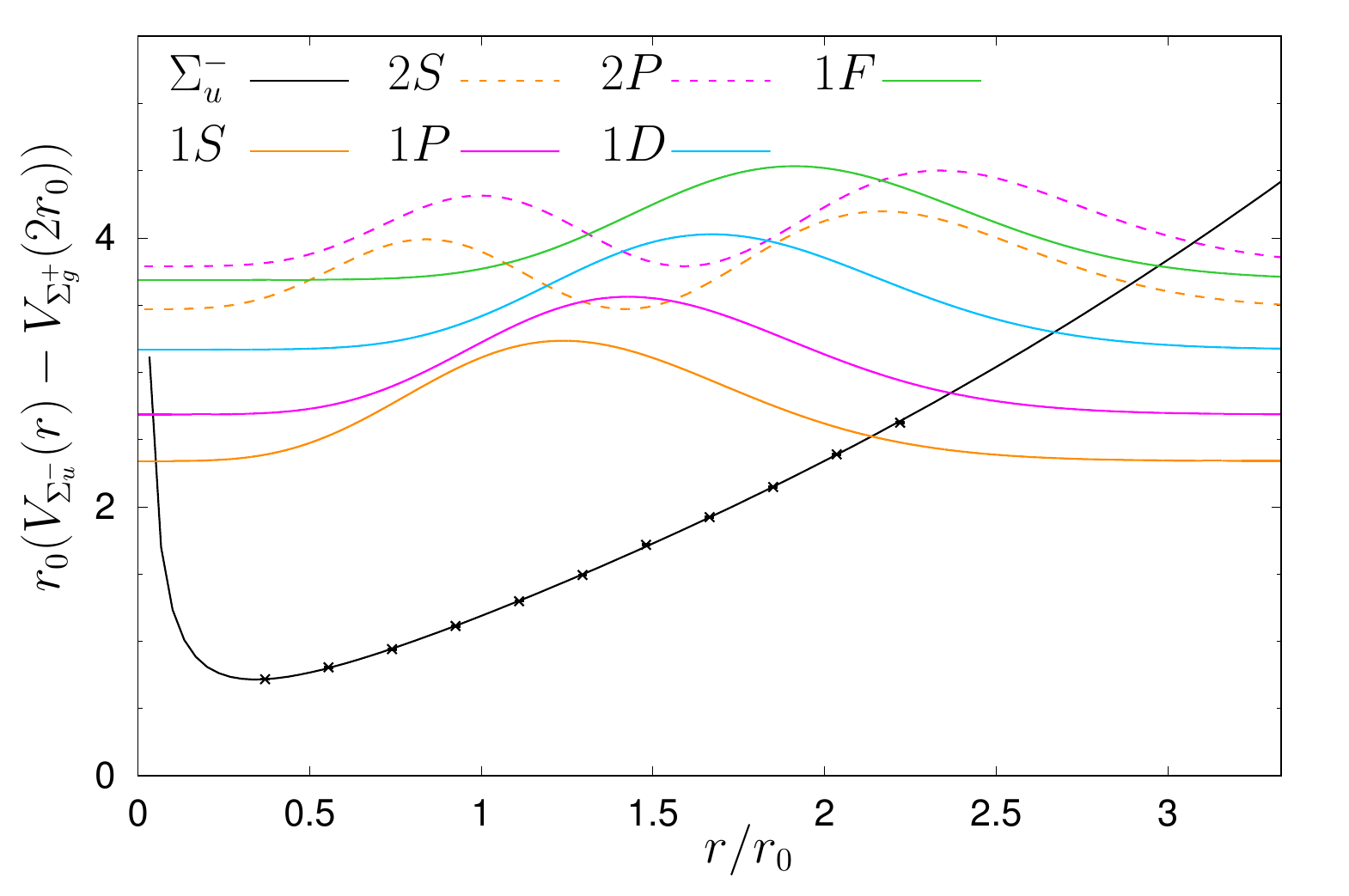}

\vspace{0.2cm}
radial probability density for $\bar{b} b$ hybrid mesons \\
\includegraphics[width=7.8cm]{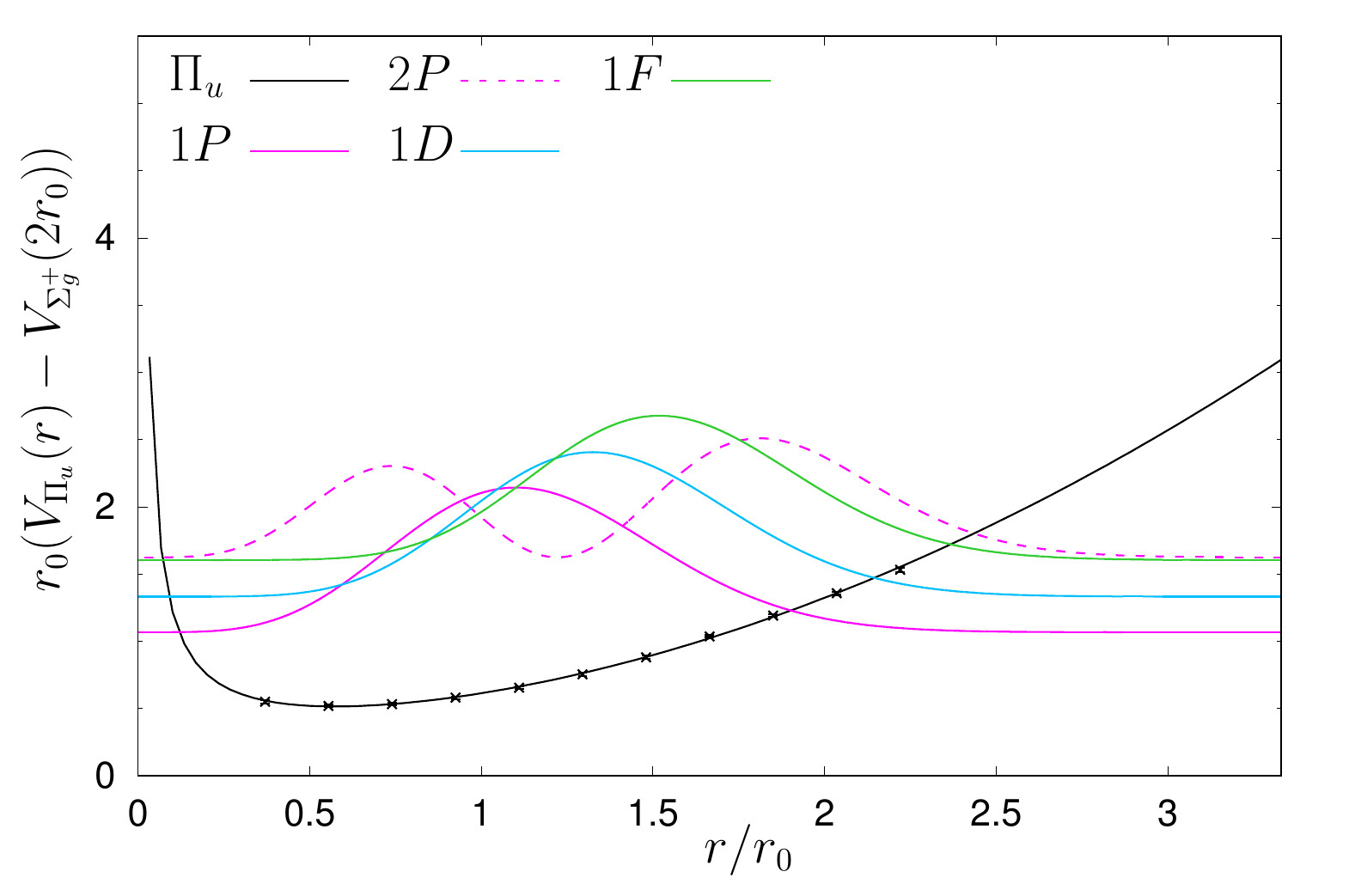}
\includegraphics[width=7.8cm]{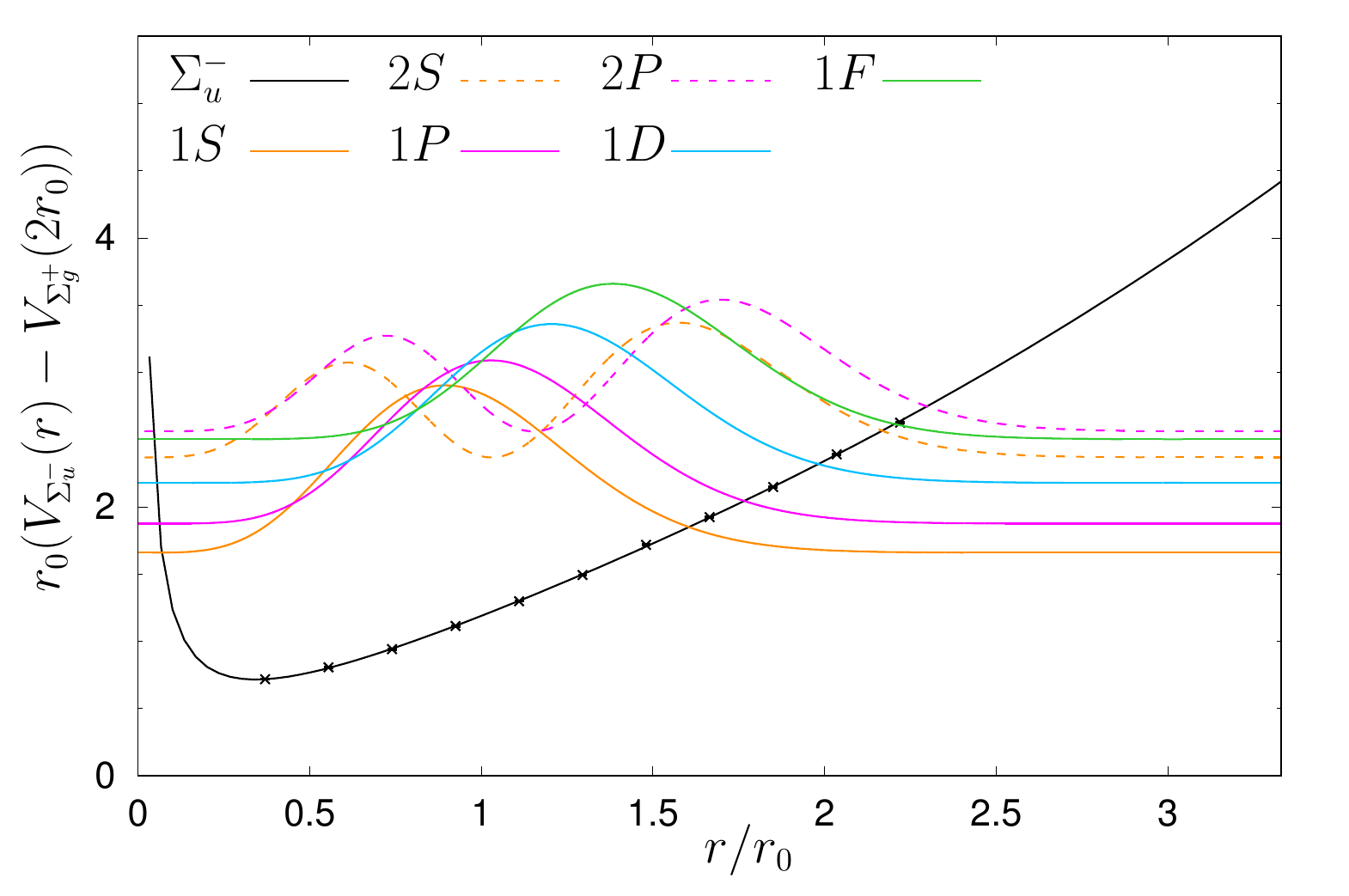}
\end{center}
\caption{\label{FIG832}Probability density for the separation $r$. \textbf{(left)}~$\Pi_u$ hybrid static potential. \textbf{(right)}~$\Sigma_u^-$ hybrid static potential.}
\end{figure}

The masses $m_{\Lambda_{\eta}^{\epsilon};L,n}$ are related to heavy hybrid mesons with quantum numbers $J^{P C}$ according to
\begin{eqnarray}
 & & \hspace{-0.7cm} J \ \ = \ \ \left\{\begin{array}{cl}
L & \textrm{if } S = 0 \\
1 & \textrm{if } S = 1 \textrm{ and } L = 0 \\
\{ L-1, L , L+1 \} & \textrm{if } S = 1 \textrm{ and } L \geq 1
\end{array}\right. \\
 & & \hspace{-0.7cm} P \ \ = \ \ \epsilon (-1)^{\Lambda + L + 1} \\
 & & \hspace{-0.7cm} C \ \ = \ \ \eta \epsilon (-1)^{\Lambda + L + S}
\end{eqnarray}
as discussed in \cite{Braaten:2014qka}. Our predicted heavy hybrid meson masses are collected in Table~\ref{TAB501} and summarized in a graphical way in Figure~\ref{FIG831}. The errors are statistical uncertainties, which have been obtained via an elaborate jackknife analysis (cf.\ section~\ref{SEC567}). In Figure~\ref{FIG832} we also show the probability density for the separation $r$, which is $|u_{\Lambda_{\eta}^{\epsilon};L,n}(r)|^2$, for the $\Pi_u$ and the $\Sigma_u^-$ hybrid static potentials and for $\bar{Q} Q = \bar{c} c$ and $\bar{Q} Q = \bar{b} b$.

\begin{table}[htb]
\begin{center}

\begin{tabular}{|c|c|c|c|c|c|c|}
\hline
 & & & & & & \vspace{-0.40cm} \\
 & & & $J^{P C}$ & $J^{P C}$ & $m_{\Lambda_{\eta}^{\epsilon};L,n}$ in MeV & $m_{\Lambda_{\eta}^{\epsilon};L,n}$ in MeV \\
$\Lambda_{\eta}^{\epsilon}$ & $L$ & $n$ & for $S=0$ & for $S=1$ & for $\bar{Q} Q = \bar{c} c$ & for $\bar{Q} Q = \bar{b} b$ \\
 & & & & & & \vspace{-0.40cm} \\
\hline
 & & & & & & \vspace{-0.40cm} \\
$\Pi_u^+$    & $1$ & $1$ & $1^{--}$ & $(0,\textbf{1},2)^{-+}$ & $4 \, 184(6)\phantom{0}$ & $10 \, 679(4)$ \\
             &     & $2$ &          &                         & $4 \, 572(10)$           & $10 \, 899(6)$ \\
             & $2$ & $1$ & $2^{++}$ & $(1,\textbf{2},3)^{+-}$ & $4 \, 374(8)\phantom{0}$ & $10 \, 783(5)$ \\
             & $3$ & $1$ & $3^{--}$ & $(2,\textbf{3},4)^{-+}$ & $4 \, 566(10)$           & $10 \, 891(6)$ \\
 & & & & & & \vspace{-0.40cm} \\
\hline
 & & & & & & \vspace{-0.40cm} \\
$\Pi_u^-$    & $1$ & $1$ & $1^{++}$ & $(\textbf{0},1,\textbf{2})^{+-}$ & $4 \, 184(6)\phantom{0}$ & $10 \, 679(4)$ \\
             &     & $2$ &          &                                  & $4 \, 572(10)$           & $10 \, 899(6)$ \\
             & $2$ & $1$ & $2^{--}$ & $(\textbf{1},2,\textbf{3})^{-+}$ & $4 \, 374(8)\phantom{0}$ & $10 \, 783(5)$ \\
             & $3$ & $1$ & $3^{++}$ & $(\textbf{2},3,\textbf{4})^{+-}$ & $4 \, 566(10)$           & $10 \, 891(6)$ \\
 & & & & & & \vspace{-0.40cm} \\
\hline
 & & & & & & \vspace{-0.40cm} \\
$\Sigma_u^-$ & $0$ & $1$ & $0^{++}$ & $1^{+-}$                & $4 \, 487(5)\phantom{0}$ & $10 \, 912(3)$ \\
             &     & $2$ &          &                         & $4 \, 933(9)\phantom{0}$ & $11 \, 192(5)$ \\
             & $1$ & $1$ & $1^{--}$ & $(0,\textbf{1},2)^{-+}$ & $4 \, 623(6)\phantom{0}$ & $10 \, 998(4)$ \\
             &     & $2$ &          &                         & $5 \, 058(10)$           & $11 \, 268(6)$ \\
             & $2$ & $1$ & $2^{++}$ & $(1,\textbf{2},3)^{+-}$ & $4 \, 814(7)\phantom{0}$ & $11 \, 117(4)$ \\
             & $3$ & $1$ & $3^{--}$ & $(2,\textbf{3},4)^{-+}$ & $5 \, 019(9)\phantom{0}$ & $11 \, 245(5)$\vspace{-0.40cm} \\
 & & & & & & \\
\hline
\end{tabular}

\caption{\label{TAB501}Predictions for heavy hybrid meson masses. Exotic $J^{P C}$ quantum numbers, i.e.\ quantum numbers forbidden in the quark model, where $P = (-1)^{L+1}$ and $C = (-1)^{L+S}$, are written in bold.}
\end{center}
\end{table}

There are also systematic errors, which are difficult to quantify. The derivation of the Schr\"odinger equation is based on several approximations, as discussed above, most notably the neglect of $1/m_Q$ corrections. In principle such corrections can be computed for hybrid static potentials, but this is expected to be very challenging, since it turned out to be difficult already for the ordinary static potential with quantum numbers $\Lambda_{\eta}^{\epsilon} = \Sigma_g^+$ \cite{Brambilla:2000gk,Pineda:2000sz,Koma:2006si,Koma:2006fw,Koma:2007jq}. At the moment we crudely estimate the magnitude of this error by half the experimental mass differences of the non-exotic $\bar{Q} Q$ mesons with $J^{P C} = 0^{-+}$ and $J^{P C} = 1^{--}$, i.e.\ $(m_{J/\Psi(1S),\textrm{exp}} - m_{\eta_c(1S),\textrm{exp}}) / 2 \approx 60 \, \textrm{MeV}$ and $(m_{\Upsilon(1S),\textrm{exp}} - m_{\eta_b(1S),\textrm{exp}}) / 2 \approx 30 \, \textrm{MeV}$. Other sources of systematic error are the finite lattice spacing and spacetime volume. We plan to improve our results in the near future by performing similar computations with even smaller values of the lattice spacing and larger spatial extent. We expect the corresponding corrections to be rather small based on our experience from previous projects concerned with the static potential, in particular \cite{Jansen:2011vv}. Moreover, dynamical light quarks have been neglected. Note, however, in \cite{Bali:2000vr} the $\Pi_u$ hybrid static potential was computed in full QCD, i.e.\ with dynamical quarks corresponding to pion masses $m_\pi \approx 500 \ldots 1200 \, \textrm{MeV}$. No statistically significant differences were found, when comparing to an equivalent computation in pure SU(3) gauge theory. We have also investigated the dependence of the predicted heavy hybrid meson masses on the $c$ and $b$ quark masses used in the Schr\"odinger equation. This dependence has been found to be very weak. For example, when using quark masses $m_c = 1480 \, \textrm{MeV}$ and $m_b = 4890 \, \textrm{MeV}$ as in \cite{Braaten:2014qka}, which differ from our choice $m_c = 1628 \, \textrm{MeV}$ and $m_b = 4977 \, \textrm{MeV}$ by $2\%$ and $9\%$, respectively, the mass differences $E_{\Lambda_{\eta}^{\epsilon};L,n} - E_{\Lambda_{\eta}^{\epsilon}=\Sigma_g^+;n=1,L=0}$ (cf.\ eq.\ (\ref{EQN864})) change for the majority of states only on the per mille level. Finally, these mass differences also depend on the scale setting procedure, i.e.\ on the value of $r_0$ used for scale setting. The corresponding relative systematic error on $E_{\Lambda_{\eta}^{\epsilon};L,n} - E_{\Lambda_{\eta}^{\epsilon}=\Sigma_g^+;n=1,L=0}$ is roughly the same as the relative uncertainty on $r_0$, which is around $4 \%$.

\newpage

$\quad$


\newpage

\section{\label{SEC602}Conclusions}

We have computed hybrid static potentials $V_{\Lambda_\eta^\epsilon}(r)$ for $\Lambda_\eta^\epsilon = \Sigma_g^-,\Sigma_u^+,\Sigma_u^-,\Pi_g,\Pi_u,\Delta_g,\Delta_u$ in SU(3) lattice gauge theory. Compared to results from the literature we use a rather fine lattice spacing and statistical errors are quite small. In contrast to the majority of existing publications, technical aspects of the computation and the analysis are discussed in detail. This offers the possibility of direct and meaningful comparison with similar computations and of methodological improvement. Moreover, we provide the numerical values of the discrete lattice data points for all computed hybrid static potentials $V_{\Lambda_\eta^\epsilon}(r)$, to allow straightforward usability of our results in future effective field theory or phenomenological work by other authors.

We also estimate masses of heavy hybrid mesons in the Born-Oppenheimer approximation, where we follow closely the approach discussed in \cite{Braaten:2014qka}. The resulting spectra are, however, rather crude estimates, mainly because $1/m_Q$ corrections, e.g.\ from the quark spins, are neglected at the moment.

An essential point of this work is the extensive optimization of hybrid static potential creation operators. We plan to use these optimized operators in follow-up projects concerned with the computation of 3-point functions. Such 3-point functions might allow to drastically reduce systematic errors in the above mentioned prediction of heavy hybrid meson masses, e.g.\ by computing quark spin corrections or by studying possible decays to ordinary quarkonium states and glueballs. Furthermore, 3-point functions will provide interesting insights concerning the gluon distribution inside heavy hybrid mesons. We have presented first corresponding results at recent conferences \cite{Mueller:2018fkg,Mueller:2018idu}.


\newpage

\appendix

\section{\label{APP123}Angular momentum of the trial states}

In this appendix we show that the trial state (\ref{EQN895}),
\begin{eqnarray}
\label{EQN642} \ket{\Psi_\text{hybrid}}_{S;\Lambda} \ \ = \ \ \int_0^{2\pi} d\varphi \, \textrm{exp}(i \Lambda \varphi) R(\varphi) O_S \ket{\Omega} ,
\end{eqnarray}
has definite total angular momentum $\Lambda$ with respect to the $z$ axis.

Any state can be rotated by an angle $\alpha$ around the $z$ axis using the total angular momentum operator $J_z$. For example the rotated trial state $\ket{\Psi_\text{hybrid}}_{S;\Lambda}$ is given by
\begin{eqnarray}
\label{EQN696} R(\alpha) \ket{\Psi_\text{hybrid}}_{S;\Lambda} \ \ = \ \ \textrm{exp}(-i J_z \alpha) \ket{\Psi_\text{hybrid}}_{S;\Lambda} .
\end{eqnarray}

The same rotation is obtained by explicitly rotating the field operators on the right hand side of eq.\ (\ref{EQN642}), which amounts to replacing the weight factor $\textrm{exp}(i \Lambda \varphi)$ by $\textrm{exp}(i \Lambda (\varphi - \alpha))$,
\begin{eqnarray}
\nonumber & & \hspace{-0.7cm} R(\alpha) \ket{\Psi_\text{hybrid}}_{S;\Lambda} \ \ = \ \ R(\alpha) \int_0^{2\pi} d\varphi \, \textrm{exp}(i \Lambda \varphi) R(\varphi) O_S \ket{\Omega} \ \ = \\
\label{EQN697} & & = \ \ \int_0^{2\pi} d\varphi \, \textrm{exp}(i \Lambda (\varphi - \alpha)) R(\varphi) O_S \ket{\Omega} \ \ = \ \ \textrm{exp}(-i \Lambda \alpha) \ket{\Psi_\text{hybrid}}_{S;\Lambda} .
\end{eqnarray}

Equating the right hand sides of eq.\ (\ref{EQN696}) and eq.\ (\ref{EQN697}) and considering an infinitesimal angle $\alpha$ leads to
\begin{eqnarray}
J_z \ket{\Psi_\text{hybrid}}_{S;\Lambda} \ \ = \ \ \Lambda \ket{\Psi_\text{hybrid}}_{S;\Lambda} .
\end{eqnarray}
This proves that the trial state $\ket{\Psi_\text{hybrid}}_{S;\Lambda}$ is an eigenstate of the total angular momentum operator $J_z$ with eigenvalue $\Lambda$.


\newpage

\section*{Acknowledgements}

We acknowledge useful discussions with Colin Morningstar.

C.R.\ acknowledges support by a Karin and Carlo Giersch Scholarship of the Giersch foundation. O.P.\ and M.W.\ acknowledge support by the DFG (German Research Foundation), grants PH 158/4-1 and WA 3000/2-1. M.W.\ acknowledges support by the Emmy Noether Programme of the DFG, grant WA 3000/1-1.

This work was supported in part by the Helmholtz International Center for FAIR within the framework of the LOEWE program launched by the State of Hesse.

Calculations on the LOEWE-CSC and on the on the FUCHS-CSC high-performance computer of the Frankfurt University were conducted for this research. We would like to thank HPC-Hessen, funded by the State Ministry of Higher Education, Research and the Arts, for programming advice.




\newpage

\begin{thebibliography}{99}

\bibitem{Braaten:2014ita} 
  E.~Braaten, C.~Langmack and D.~H.~Smith,
  ``Selection rules for hadronic transitions of XYZ mesons,''
  Phys.\ Rev.\ Lett.\ {\bf 112}, 222001 (2014)
  [arXiv:1401.7351 [hep-ph]].

\bibitem{Meyer:2015eta} 
  C.~A.~Meyer and E.~S.~Swanson,
  ``Hybrid mesons,''
  Prog.\ Part.\ Nucl.\ Phys.\ {\bf 82}, 21 (2015)
  [arXiv:1502.07276 [hep-ph]].

\bibitem{Swanson:2015wgq} 
  E.~S.~Swanson,
  ``$XYZ$ states: theory overview,''
  AIP Conf.\ Proc.\ {\bf 1735}, 020013 (2016)
  [arXiv:1512.04853 [hep-ph]].

\bibitem{Lebed:2016hpi} 
  R.~F.~Lebed, R.~E.~Mitchell and E.~S.~Swanson,
  ``Heavy-quark QCD exotica,''
  Prog.\ Part.\ Nucl.\ Phys.\ {\bf 93}, 143 (2017)
  [arXiv:1610.04528 [hep-ph]].

\bibitem{Olsen:2017bmm} 
  S.~L.~Olsen, T.~Skwarnicki and D.~Zieminska,
  ``Nonstandard heavy mesons and baryons: experimental evidence,''
  Rev.\ Mod.\ Phys.\  {\bf 90}, 015003 (2018)
  [arXiv:1708.04012 [hep-ph]].

\bibitem{Braaten:2014qka} 
  E.~Braaten, C.~Langmack and D.~H.~Smith,
  ``Born-Oppenheimer approximation for the $XYZ$ Mesons,''
  Phys.\ Rev.\ D {\bf 90}, 014044 (2014)
  [arXiv:1402.0438 [hep-ph]].

\bibitem{Berwein:2015vca} 
  M.~Berwein, N.~Brambilla, J.~Tarrus Castella and A.~Vairo,
  ``Quarkonium hybrids with nonrelativistic effective field theories,''
  Phys.\ Rev.\ D {\bf 92}, 114019 (2015)
  [arXiv:1510.04299 [hep-ph]].

\bibitem{Griffiths:1983ah} 
  L.~A.~Griffiths, C.~Michael and P.~E.~L.~Rakow,
  ``Mesons with excited glue,''
  Phys.\ Lett.\ {\bf 129B}, 351 (1983).

\bibitem{Campbell:1984fe} 
  N.~A.~Campbell, L.~A.~Griffiths, C.~Michael and P.~E.~L.~Rakow,
  ``Mesons with excited glue from SU(3) lattice gauge theory,''
  Phys.\ Lett.\ {\bf 142B}, 291 (1984).

\bibitem{Campbell:1987nv} 
  N.~A.~Campbell, A.~Huntley and C.~Michael,
  ``Heavy quark potentials and hybrid mesons from SU(3) lattice gauge theory,''
  Nucl.\ Phys.\ B {\bf 306}, 51 (1988).

\bibitem{Michael:1990az} 
  C.~Michael and S.~J.~Perantonis,
  ``Potentials and glueballs at large beta in SU(2) pure gauge theory,''
  J.\ Phys.\ G {\bf 18}, 1725 (1992).

\bibitem{Perantonis:1990dy} 
  S.~Perantonis and C.~Michael,
  ``Static potentials and hybrid mesons from pure SU(3) lattice gauge theory,''
  Nucl.\ Phys.\ B {\bf 347}, 854 (1990).

\bibitem{Juge:1997nc} 
  K.~J.~Juge, J.~Kuti and C.~J.~Morningstar,
  ``Gluon excitations of the static quark potential and the hybrid quarkonium spectrum,''
  Nucl.\ Phys.\ Proc.\ Suppl.\ {\bf 63}, 326 (1998)
  [hep-lat/9709131].

\bibitem{Peardon:1997jr} 
  M.~J.~Peardon,
  ``Coarse lattice results for glueballs and hybrids,''
  Nucl.\ Phys.\ Proc.\ Suppl.\ {\bf 63}, 22 (1998)
  [hep-lat/9710029].

\bibitem{Juge:1997ir} 
  K.~J.~Juge, J.~Kuti and C.~J.~Morningstar,
  ``A study of hybrid quarkonium using lattice QCD,''
  AIP Conf.\ Proc.\ {\bf 432}, 136 (1998)
  [hep-ph/9711451].

\bibitem{Morningstar:1998xh} 
  C.~Morningstar, K.~J.~Juge and J.~Kuti,
  ``Gluon excitations of the static quark potential,''
  hep-lat/9809015.

\bibitem{Michael:1998tr} 
  C.~Michael,
  ``Hadronic spectroscopy from the lattice: glueballs and hybrid mesons,''
  Nucl.\ Phys.\ A {\bf 655}, 12 (1999)
  [hep-ph/9810415].

\bibitem{Juge:1999ie} 
  K.~J.~Juge, J.~Kuti and C.~J.~Morningstar,
  ``Ab initio study of hybrid $\bar{b} g b$ mesons,''
  Phys.\ Rev.\ Lett.\ {\bf 82}, 4400 (1999)
  [hep-ph/9902336].

\bibitem{Juge:1999aw} 
  K.~J.~Juge, J.~Kuti and C.~J.~Morningstar,
  ``The heavy hybrid spectrum from NRQCD and the Born-Oppenheimer approximation,''
  Nucl.\ Phys.\ Proc.\ Suppl.\ {\bf 83}, 304 (2000)
  [hep-lat/9909165].

\bibitem{Michael:1999ge} 
  C.~Michael,
  ``Quarkonia and hybrids from the lattice,''
  PoS HF {\bf 8}, 001 (1999)
  [hep-ph/9911219].

\bibitem{Bali:2000vr} 
  G.~S.~Bali {\it et al.} [SESAM and T$\chi$L Collaborations],
  ``Static potentials and glueball masses from QCD simulations with Wilson sea quarks,''
  Phys.\ Rev.\ D {\bf 62}, 054503 (2000)
  [hep-lat/0003012].

\bibitem{Morningstar:2001nu} 
  C.~Morningstar,
  ``Gluonic excitations in lattice QCD: a brief survey,''
  AIP Conf.\ Proc.\ {\bf 619}, 231 (2002)
  [nucl-th/0110074].

\bibitem{Juge:2002br} 
  K.~J.~Juge, J.~Kuti and C.~Morningstar,
  ``Fine structure of the QCD string spectrum,''
  Phys.\ Rev.\ Lett.\ {\bf 90}, 161601 (2003)
  [hep-lat/0207004].

\bibitem{Michael:2003ai} 
  C.~Michael,
  ``Exotics,''
  Int.\ Rev.\ Nucl.\ Phys.\ {\bf 9}, 103 (2004)
  [hep-lat/0302001].

\bibitem{Juge:2003qd} 
  K.~J.~Juge, J.~Kuti and C.~Morningstar,
  ``The Heavy quark hybrid meson spectrum in lattice QCD,''
  AIP Conf.\ Proc.\ {\bf 688}, 193 (2004)
  [nucl-th/0307116].

\bibitem{Michael:2003xg} 
  C.~Michael,
  ``Hybrid mesons from the lattice,''
  hep-ph/0308293.

\bibitem{Bali:2003jq} 
  G.~S.~Bali and A.~Pineda,
  ``QCD phenomenology of static sources and gluonic excitations at short distances,''
  Phys.\ Rev.\ D {\bf 69}, 094001 (2004)
  [hep-ph/0310130].

\bibitem{Juge:2003ge} 
  K.~J.~Juge, J.~Kuti and C.~Morningstar,
  ``Excitations of the static quark anti-quark system in several gauge theories,''
  hep-lat/0312019.

\bibitem{Wolf:2014tta} 
  P.~Wolf and M.~Wagner,
  ``Lattice study of hybrid static potentials,''
  J.\ Phys.\ Conf.\ Ser.\ {\bf 599}, 012005 (2015)
  [arXiv:1410.7578 [hep-lat]].

\bibitem{Reisinger:2017btr} 
  C.~Reisinger, S.~Capitani, O.~Philipsen and M.~Wagner,
  ``Computation of hybrid static potentials in SU(3) lattice gauge theory,''
  EPJ Web Conf.\ {\bf 175}, 05012 (2018)
  [arXiv:1708.05562 [hep-lat]].

\bibitem{Bicudo:2018yhk} 
  P.~Bicudo, M.~Cardoso and N.~Cardoso,
  ``Colour fields of the quark-antiquark excited flux tube,''
  EPJ Web Conf.\ {\bf 175}, 14009 (2018)
  [arXiv:1803.04569 [hep-lat]].

\bibitem{Bicudo:2018jbb} 
  P.~Bicudo, N.~Cardoso and M.~Cardoso,
  ``Colour field densities of the quark-antiquark excited flux tubes in SU(3) lattice QCD,''
  arXiv:1808.08815 [hep-lat].

\bibitem{Reisinger:2018lne} 
  C.~Reisinger, S.~Capitani, L.~M\"uller, O.~Philipsen and M.~Wagner,
  ``Computation of hybrid static potentials from optimized trial states in SU(3) lattice gauge theory,''
  arXiv:1810.13284 [hep-lat].

\bibitem{bo}
  M.~Born and J.~R.~Oppenheimer,
  ``Zur Quantentheorie der Molekeln'',
  Annalen der Physik {\bf 389} (20), 457–484 (1927).

\bibitem{Brambilla:2017uyf} 
  N.~Brambilla, G.~Krein, J.~Tarrus Castella and A.~Vairo,
  ``Born-Oppenheimer approximation in an effective field theory language,''
  Phys.\ Rev.\ D {\bf 97}, 016016 (2018)
  [arXiv:1707.09647 [hep-ph]].

\bibitem{Brambilla:2018pyn} 
  N.~Brambilla, W.~K.~Lai, J.~Segovia, J.~Tarruscasteli and A.~Vairo,
  ``Spin structure of heavy-quark hybrids,''
  arXiv:1805.07713 [hep-ph].

\bibitem{Bali:2005fu} 
  G.~S.~Bali {\it et al.} [SESAM Collaboration],
  ``Observation of string breaking in QCD,''
  Phys.\ Rev.\ D {\bf 71}, 114513 (2005)
  [hep-lat/0505012].

\bibitem{Bicudo:2015kna} 
  P.~Bicudo, K.~Cichy, A.~Peters and M.~Wagner,
  ``$B B$ interactions with static bottom quarks from Lattice QCD,''
  Phys.\ Rev.\ D {\bf 93}, 034501 (2016)
  [arXiv:1510.03441 [hep-lat]].

\bibitem{Cornwell:1997ke} 
  J.~F.~Cornwell,
  ``Group theory in physics: an introduction,''
  Academic Press (1997).

\bibitem{Rothe:1992nt} 
  H.~J.~Rothe,
  ``Lattice gauge theories: an Introduction,''
  World Sci.\ Lect.\ Notes Phys.\ {\bf 82}, 1 (2012).

\bibitem{Edwards:2004sx} 
  R.~G.~Edwards {\it et al.} [SciDAC and LHPC and UKQCD Collaborations],
  ``The Chroma software system for lattice QCD,''
  Nucl.\ Phys.\ Proc.\ Suppl.\  {\bf 140}, 832 (2005)
  [hep-lat/0409003].

\bibitem{Hasenfratz:2001hp}
  A.~Hasenfratz and F.~Knechtli,
  ``Flavour symmetry and the static potential with hypercubic blocking,''
  Phys.\ Rev.\ D {\bf 64}, 034504 (2001)
  [arXiv:hep-lat/0103029].

\bibitem{DellaMorte:2003mn}
  M.~Della Morte {\it et al.},
  ``Lattice HQET with exponentially improved statistical precision,''
  Phys. Lett. {\bf B581}, 93, (2004)
  [arXiv:hep-lat/0307021].

\bibitem{Della Morte:2005yc}
  M.~Della Morte, A.~Shindler and R.~Sommer,
  ``On lattice actions for static quarks,''
  JHEP {\bf 0508}, 051 (2005)
  [arXiv:hep-lat/0506008].

\bibitem{Jansen:2008si} 
  K.~Jansen {\it et al.} [ETM Collaboration],
  ``The Static-light meson spectrum from twisted mass lattice QCD,''
  JHEP {\bf 0812}, 058 (2008)
  [arXiv:0810.1843 [hep-lat]].

\bibitem{Luscher:1990ck} 
  M.~L\"uscher and U.~Wolff,
  ``How to calculate the elastic scattering matrix in two-dimensional quantum field theories by numerical simulation,''
  Nucl.\ Phys.\ B {\bf 339}, 222 (1990).

\bibitem{Blossier:2009kd} 
  B.~Blossier, M.~Della Morte, G.~von Hippel, T.~Mendes and R.~Sommer,
  ``On the generalized eigenvalue method for energies and matrix elements in lattice field theory,''
  JHEP {\bf 0904}, 094 (2009)
  [arXiv:0902.1265 [hep-lat]].

\bibitem{Oncala:2017hop} 
  R.~Oncala and J.~Soto,
  ``Heavy quarkonium hybrids: spectrum, decay and mixing,''
  Phys.\ Rev.\ D {\bf 96}, 014004 (2017)
  [arXiv:1702.03900 [hep-ph]].

\bibitem{wwwCM}
  \texttt{http://www.andrew.cmu.edu/user/cmorning/static\_potentials/SU3\_4D/greet.html}.

\bibitem{Brambilla:1999xf} 
  N.~Brambilla, A.~Pineda, J.~Soto and A.~Vairo,
  ``Potential NRQCD: an effective theory for heavy quarkonium,''
  Nucl.\ Phys.\ B {\bf 566}, 275 (2000)
  [hep-ph/9907240].

\bibitem{Morningstar:1999rf} 
  C.~J.~Morningstar and M.~J.~Peardon,
  ``The Glueball spectrum from an anisotropic lattice study,''
  Phys.\ Rev.\ D {\bf 60}, 034509 (1999)
  [hep-lat/9901004].

\bibitem{Karbstein:2018mzo} 
  F.~Karbstein, M.~Wagner and M.~Weber,
  ``Determination of $\Lambda_{\overline{\textrm{MS}}}^{(n_f=2)}$ and analytic parameterization of the static quark-antiquark potential,''
  Phys.\ Rev.\ D {\bf 98}, 114506 (2018)
  [arXiv:1804.10909 [hep-ph]].

\bibitem{Sommer:2014mea} 
  R.~Sommer,
  ``Scale setting in lattice QCD,''
  PoS LATTICE {\bf 2013}, 015 (2014)
  [arXiv:1401.3270 [hep-lat]].

\bibitem{Koma:2006si} 
  Y.~Koma, M.~Koma and H.~Wittig,
  ``Nonperturbative determination of the QCD potential at $\mathcal{O}(1/m)$,''
  Phys.\ Rev.\ Lett.\ {\bf 97}, 122003 (2006)
  [hep-lat/0607009].

\bibitem{Hasenfratz:1980jv} 
  P.~Hasenfratz, R.~R.~Horgan, J.~Kuti and J.~M.~Richard,
  ``The effects of colored glue in the QCD motivated bag of heavy quark -- anti-quark systems,''
  Phys.\ Lett.\ {\bf 95B}, 299 (1980).

\bibitem{Godfrey:1985xj} 
  S.~Godfrey and N.~Isgur,
  ``Mesons in a relativized quark model with chromodynamics,''
  Phys.\ Rev.\ D {\bf 32}, 189 (1985).

\bibitem{Tanabashi:2018oca} 
  M.~Tanabashi {\it et al.} [Particle Data Group],
  ``Review of particle physics,''
  Phys.\ Rev.\ D {\bf 98}, 030001 (2018).

\bibitem{Brambilla:2000gk} 
  N.~Brambilla, A.~Pineda, J.~Soto and A.~Vairo,
  ``The QCD potential at $\mathcal{O}(1/m)$,''
  Phys.\ Rev.\ D {\bf 63}, 014023 (2001)
  [hep-ph/0002250].

\bibitem{Pineda:2000sz} 
  A.~Pineda and A.~Vairo,
  ``The QCD potential at $\mathcal{O}(1/m^2)$: complete spin dependent and spin independent result,''
  Phys.\ Rev.\ D {\bf 63}, 054007 (2001)
  [Erratum: Phys.\ Rev.\ D {\bf 64}, 039902 (2001)]
  [hep-ph/0009145].

\bibitem{Koma:2006fw} 
  Y.~Koma and M.~Koma,
  ``Spin-dependent potentials from lattice QCD,''
  Nucl.\ Phys.\ B {\bf 769}, 79 (2007)
  [hep-lat/0609078].

\bibitem{Koma:2007jq} 
  Y.~Koma, M.~Koma and H.~Wittig,
  ``Relativistic corrections to the static potential at $\mathcal{O}(1/m)$ and $\mathcal{O}(1/m^2)$,''
  PoS LATTICE {\bf 2007}, 111 (2007)
  [arXiv:0711.2322 [hep-lat]].

\bibitem{Jansen:2011vv} 
  K.~Jansen {\it et al.} [ETM Collaboration],
  ``$\Lambda_{\overline{\textrm{MS}}}$ from the static potential for QCD with $n_f=2$ dynamical quark flavors,''
  JHEP {\bf 1201}, 025 (2012)
  [arXiv:1110.6859 [hep-ph]].

\bibitem{Mueller:2018fkg} 
  L.~M\"uller and M.~Wagner,
  ``Structure of hybrid static potential flux tubes in SU(2) lattice Yang-Mills theory,''
  Acta Phys.\ Polon.\ Supp.\  {\bf 11}, 551 (2018)
  [arXiv:1803.11124 [hep-lat]].

\bibitem{Mueller:2018idu} 
  L.~M\"uller, O.~Philipsen, C.~Reisinger and M.~Wagner,
  ``Structure of hybrid static potential flux tubes in lattice Yang-Mills theory,''
  arXiv:1811.00452 [hep-lat].

\end{thebibliography}
\end{document}